\documentclass[lettersize,journal]{IEEEtran}
\usepackage{amsmath,amssymb,amsfonts}
\usepackage{algorithmic}
\usepackage{algorithm}
\usepackage{array}
\usepackage[caption=false,font=footnotesize,labelfont=sf,textfont=sf]{subfig}
\usepackage{textcomp}
\usepackage{stfloats}
\usepackage{url}
\usepackage{xcolor}
\usepackage{verbatim}
\usepackage{graphicx}
\usepackage{cite}

\usepackage{multicol}
\usepackage{pifont}
\usepackage{soul}
\usepackage{multirow}
\usepackage{booktabs}
\usepackage{tablefootnote}
\usepackage{tabularx}
\usepackage{soul}

\begin{document}

\title{Allspark: Workload Orchestration for Visual Transformers on Processing In-Memory Systems}

\author{Mengke Ge, Junpeng Wang, Binhan Chen, Yingjian Zhong, Haitao Du, \\Song Chen,~\IEEEmembership{Member,~IEEE,} and Yi Kang,~\IEEEmembership{Member,~IEEE}
        % <-this % stops a space
\thanks{Manuscript received April 19, 2021; revised August 16, 2021. (\emph{Corresponding authors: Song Chen and Yi Kang}) }
\thanks{Mengke Ge, Song Chen, and Yi Kang are with Institute of Artificial Intelligence, Hefei Comprehensive National Science Center, Hefei, China.}
\thanks{Junpeng Wang, Binhan Chen, Haitao Du, Song Chen, and Yi Kang are with School of Microelectronics, University of Science and Technology of China (USTC), Hefei, China.}
\thanks{Yingjian Zhong is with Anhui University, Hefei, China.}
\thanks{Email: \texttt{mengke.ge@iai.ustc.edu.cn, songch@ustc.edu.cn, ykang@ustc.edu.cn.}}}

\markboth{Journal of \LaTeX\ Class Files,~Vol.~14, No.~8, August~2021}%
{Shell \MakeLowercase{\textit{et al.}}: A Sample Article Using IEEEtran.cls for IEEE Journals}

\IEEEpubid{0000--0000/00\$00.00~\copyright~2021 IEEE}

\maketitle

\begin{abstract}
The advent of Transformers has revolutionized computer vision, offering a powerful alternative to convolutional neural networks (CNNs), especially with the local attention mechanism that excels at capturing local structures within the input and achieve state-of-the-art performance.
Processing in-memory (PIM) architecture offers extensive parallelism, low data movement costs, and scalable memory bandwidth, making it a promising solution to accelerate Transformer with memory-intensive operations.
However, the crucial issue lies in efficiently deploying an entire model onto resource-limited PIM system while parallelizing each transformer block with potentially many computational branches based on local-attention mechanisms.

We present Allspark, which focuses on workload orchestration for visual Transformers on PIM systems, aiming at minimizing inference latency.
Firstly, to fully utilize the massive parallelism of PIM, Allspark employs a fine-grained partitioning scheme for computational branches, and formats a systematic layout and interleaved dataflow with maximized data locality and reduced data movement.
Secondly, Allspark formulates the scheduling of the complete model on a resource-limited distributed PIM system as an integer linear programming (ILP) problem.
Thirdly, as local-global data interactions exhibit complex yet regular dependencies, Allspark provides a two-stage placement method, which simplifies the challenging placement of computational branches on the PIM system into the structured layout and greedy-based binding, to minimize NoC communication costs.
Extensive experiments on 3D-stacked DRAM-based PIM systems show that Allspark brings $1.2\times$$\sim$$24.0\times$ inference speedup for various visual Transformers over baselines.
Compared to Nvidia V100 GPU, Allspark-enriched PIM system yields average speedups of $2.3\times$ and energy savings of $20\times$$\sim$$55\times$.
\end{abstract}


\section{Introduction}
\label{sec:intro}
\IEEEPARstart{T}{ransformer}, an attention-based neural network, has attracted tremendous interests due to their effectiveness in various domains such as language, computer vision (CV), and reinforcement learning\cite{Attention,TPAMI}.
Visual Transformers, such as ViT\cite{ViT} and PVT\cite{PVT}, have shown impressive performance in CV tasks such as image classification, object detection, and semantic segmentation, even outperforming the go-to architecture CNNs, thanks to their larger receptive fields capable of capturing long-range dependencies between patches\cite{TPAMI}.
Recently, local-attention visual Transformers (LVTs)\cite{Swin,Focal,Twins,shuffle,CSWin} have enjoyed great popularity. By virtue of their adeptness at effectively capturing the local structure in the input, LVTs have shown noteworthy improvements in performance compared to original visual Transformers, and are well ahead of the leaderboards of various CV datasets\cite{sota}.

Unfortunately, Transformers run significantly sluggishly on general-purpose platforms such as GPU and CPU.
Due to the high memory footprint, low data reuse, and complex data movement of operations such as self-attention, Transformers exhibit memory-intensive characteristics \cite{TransPIM,FACT}.
Thus, memory bandwidth becomes a crucial bottleneck, resulting in under-utilization of computing units and low arithmetic intensity (FLOPs/Byte).
To accelerate inference, Transformer accelerators\cite{ViTCoD,Spatten,DOTA,FACT} have been developed to offload either self-attention operation or the entire model from general-purpose platforms.
However, these accelerators necessitate loading data from off-chip memory, and parallelism and bandwidth remain insufficient, which limits acceleration performance.

Processing in-memory (PIM) or Processing near-memory architectures have the ability to effectively alleviate the memory wall problem by moving the computations closer to the data locations in the main memory, and have been used to accelerate memory-intensive applications\cite{GraphH,Neurocube,Newton,VLSI,TETRIS,AiM,3DDRAM}.
For DNN acceleration, PIM architectures typically employ a spatial structure (tiled structure)\cite{VLSI,Neurocube,TETRIS,DOJO,nicepim}, which is a scalable 2D array of compute nodes connected via network-on-chip (NoC).
Each node \textbf{(PIM-node)} comprises a memory subsystem (typically DRAM\cite{VLSI,Neurocube,TETRIS} or SRAM\cite{DOJO}) and a processing engine.
With its extensive parallelism, low data movement cost, and high memory bandwidth, PIM architecture has reaped more attention to accelerate Transformers, and recent studies\cite{TransPIM,MAT} have shown that PIM architectures enabled Transformers to achieve significant improvements in both inference speed and energy efficiency over GPUs.

The deployment of visual Transformers is a significant challenge, especially with the introduction of a local attention mechanism.
Original Transformers employ global attention whose computational complexity is quadratic over all the input\cite{TPAMI}, and the attention computation acts as a bottleneck when dealing with dense visual inputs, such as pixel-level semantic segmentation of high-resolution images.
To reduce the computing complexity and memory demands, LVTs are born, which introduce a local attention mechanism.
In addition, LVTs also exhibit superior expressiveness and generalization\cite{TPAMI}.
All image patches are divided into many small-sized local regions (subsets of adjacent patches), each of which independently serves as an input to attention computation.
These local regions typically have a fixed shape such as a window\cite{Swin,Focal,Twins} and block pattern of fixed strides\cite{CSWin}.
Visual Transformers consist of many consecutive transformer blocks, whilst LVTs have many local-region based attention computations (called \emph{\textbf{computational branches}}) on each transformer block, allowing high computational parallelism.
Moreover, visual Transformers exhibit a multi-stage hierarchical structure (Section \ref{sec:back_trans}), causing variations in the number of computational branches and computation on transformer blocks at different stages.

Therefore, the crucial issue lies in efficiently deploying an entire visual Transformers onto the PIM system while parallelizing each transformer block, with potentially many local-attention-based computational branches.
Besides, visual Transformers up to hundreds of millions of parameters are computing and memory demanding, making deployment in the PIM system with limited and distributed resources challenging.
Previous works\cite{ViTCoD,Spatten,DOTA,FACT,TransPIM} exploit a fixed and uni-patterned parallelism tailored to global attention, but these coarse-grained partitioning yields insufficient parallelism.
Existing DNN mappers\cite{TANGRAM,Timeloop,Atomic,CoSA,inter-layer-scheduling} target architectures with a multi-level memory hierarchy, and they do not cover data placement across on-chip distributed memory. It has been shown that these mappers incur high data movement costs\cite{TransPIM}.
To bridge the gap, we propose a deployment framework for visual Transformers, which efficiently enables end-to-end inference on PIM systems without off-chip memory. Our contributions are outlined below:

\begin{itemize}
\item Allspark endeavors to orchestrate the workload for visual Transformers on PIM systems, aiming at minimizing inference latency. To the best of our knowledge, Allspark is the first deployment framework dedicated to accelerating visual Transformers inference on PIM systems, providing a new perspective on efficient PIM solutions.

\item To fully utilize the massive parallelism of PIM, Allspark employs a fine-grained partitioning for computational branches, and format a systematic layout and interleaved dataflows with maximized data locality and reduced frequent data movement between PIM-nodes.

\item Allspark formulates the scheduling of the complete model on a resource-limited distributed PIM system as an ILP problem, and all computational branches are temporally and spatially scheduled to maximize resource utilization.

\item As local-global data interactions exhibit complex yet regular dependencies, Allspark offers a two-stage method, which simplifies the challenging placement of computational branches on the PIM system into the structured layout and greedy-based binding, to minimize NoC communication costs.

\item Extensive experiments on DRAM-based PIM systems show that Allspark brings 1.2$\times$$\sim$24.0$\times$ inference speedup over baselines, and Allspark-enriched PIM system yields average speedups of 2.3$\times$ and energy savings of $20\times$$\sim$$55\times$ over Nvidia V100 GPU.
\end{itemize}

\section{Related Work}

\subsection{Transformer Accelerators}

Recently, a plethora of ASIC-based Transformer accelerators have emerged\cite{ViTCoD,Spatten,DOTA,FACT}.
These accelerators employ a software-architecture co-design approach, aiming to accelerate the execution of self-attention mechanisms, or the inference of the entire NLP Transformer and the original vision Transformer, both of which rely on the conventional global attention mechanism.
Table \ref{table:SOTA_Arch} provides a summary of these advanced accelerators.
However, ASIC-based accelerators face a bottleneck in off-chip memory bandwidth, particularly for memory-intensive layers in Transformers. In contrast, PIM systems offer a solution by storing all data in memory, eliminating the need for costly off-chip data transfers.

TransPIM\cite{TransPIM} is a software-hardware co-design solution based on the PIM architecture for inference acceleration in NLP Transformers.
TransPIM introduces token-based dataflow and lightweight modifications into the high bandwidth memory (HBM) architecture to support computation and memory operations in Transformers.
MAT\cite{MAT}, a PIM framework to accelerate long-sequence attention, adopts a memory-efficient processing flow to process sub-sequences in a pipeline with a small memory footprint, yielding significant improvements in speed and energy consumption.
However, MAT is only targeted at the attention layer.
In short, PIM-based architectures bring notable acceleration performance gains in both NLP and CV domains for Transformers based on global attention.

\begin{table*}[htbp]
  \centering
  \fontsize{7.5}{6}\selectfont
  \caption{Summary of SOTA Transformer Accelerators.}
  \label{table:SOTA_Arch}
  \begin{tabular}{llllc}
    \toprule
     & \textbf{Architecture}& \textbf{Processing Unit} & \textbf{Model} & \textbf{Partitioning \& Mapping Granularity} \\
    \midrule
    \textbf{SpAtten} \cite{Spatten}& Dedicated Arch.+Model Pruning& Vector-Matrix Mult.& BERT, GPT-2  & \color{black}{patch-level}\\
    \textbf{TransPIM} \cite{TransPIM}& Dedicated PIM-based Accelerator& Vector Mult. & RoBERTa, Pegasus, GPT-2 &  \color{black}{patch-level}\\
    \textbf{MAT} \cite{MAT}& DRAM-based PIM& - & BERT, ViT &  \color{black}{patch-level}\\
    \textbf{FACT} \cite{FACT}& Dedicated Arch.+Model Pruning& Matrix Mult. &BERT, ViT &  \color{black}{patch-level}\\
    \textbf{DOTA} \cite{DOTA}& Dedicated Arch.+Model Pruning& Reconfigurable Matrix Mult. & BERT, GPT-2 &  \color{black}{attention-head-level}\\
    \textbf{ViTCoD} \cite{ViTCoD}& Dedicated Arch.+Model Pruning& Matrix Mult.& DeiT, LeViT &  \color{black}{attention-head-level}\\
    \textbf{Vanilla} & PIM & Matrix Mult. & ViTs (even LVTs) & \color{black}{branch-level} \\
    \textbf{Allspark} (Ours)&PIM& Matrix Mult.& ViTs (even LVTs) &  \textbf{flexible fine-grained partitioning}\\
    \bottomrule
  \end{tabular}
\end{table*}

\subsection{Scheduling and Mapping Space Exploration}
\label{sec:relatedwork}

Extensive studies\cite{TANGRAM,Timeloop,Atomic,CoSA,inter-layer-scheduling} have addressed the problem of mapping DNNs to scalable tiled accelerators with a multi-level memory hierarchy.
For traditional DNN accelerators, all weights are loaded from an off-chip DRAM and then intermediate data is frequently read/written back at runtime, and loop tiling, ordering, and spatial mapping are determined at a layer granularity.
These methods are broadly categorized into \emph{inter-layer pipeline parallelism}\cite{TANGRAM,Atomic,Neurocube}, which orchestrates all layers in a pipeline manner, and \emph{operator parallelism}\cite{Timeloop,CoSA}, which dedicates all computing resources to processing each layer.
Nonetheless, for PIM architectures with distributed characteristics, these methods lead to inefficient utilization of on-chip distributed computing and memory resources\cite{DDAM,nicepim,TransPIM}.
Moreover, previous researches\cite{TransPIM} have proven that these methods are not optimally suited for Transformer models due to these layer-based parallelisms incurring significant non-computational overheads, necessitating the transfer of a large amount of inputs and weights between layers.

As shown in Table \ref{table:SOTA_Arch}, emerging accelerators adopt either attention-head-level (\textbf{AH})\cite{DOTA,ViTCoD} or patch(token)-level (\textbf{P})\cite{MAT,TransPIM,Spatten,FACT} paralleling, respectively, to accelerate inference for Transformers based on global attention.
The model is simply partitioned along attention heads, or the long sequence of inputs are partitioned into several equal fractions for parallel processing.
However, LVTs are characterized by many computational branches and even hierarchical representations (see Section \ref{sec:back_trans}), thus these fixed and uni-patterned approaches to parallelism are not efficient when deploying visual Transformers on PIM systems with distributed characteristics.
Crucially, model parallelism must contemplate efficient allocation of on-chip limited computational and memory resources, which has not been covered by existing researches.

\section{Background and Motivation}
\subsection{Visual Transformers}
\label{sec:back_trans}

\begin{figure*}[tbhp]
\centering
\includegraphics[width=0.8\textwidth]{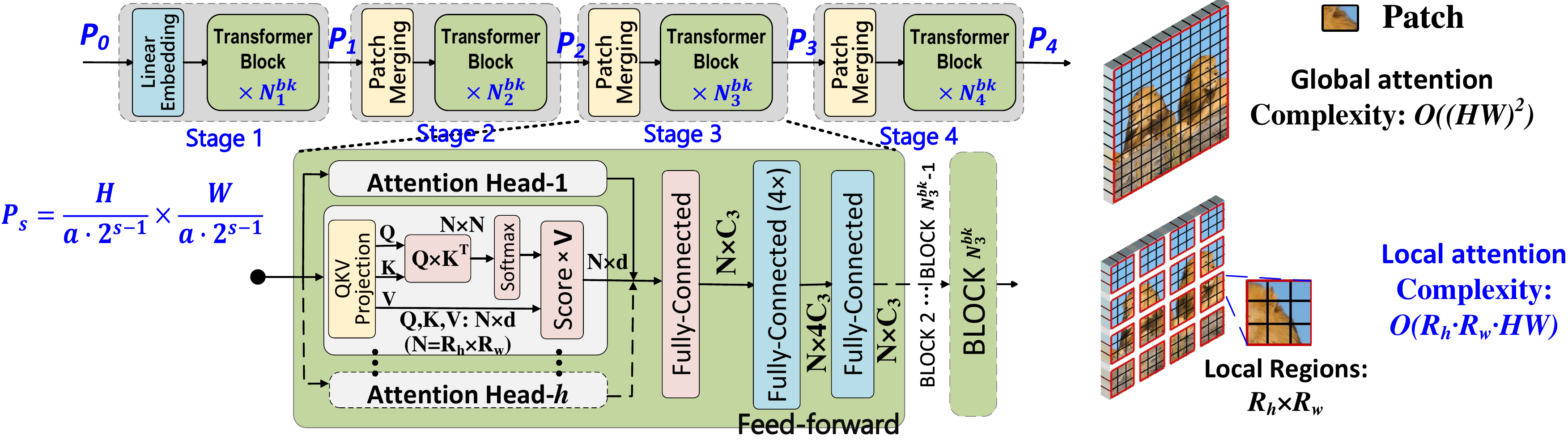}
\caption{Visual Transformer structure.}
\label{fig:Transformer}
\end{figure*}

\textbf{Structure:}
Visual Transformer generally consists of many successive transformer blocks with a multi-stage hierarchical representation as shown in Figure \ref{fig:Transformer}.
Most models are divided into $S$ (typically 4) stages, each stage $s\in[1,S]$ contains $N^{bk}_s$ sequential \emph{transformer blocks} (encoders) that extract feature representations.
Moreover, cutting-edge models (e.g. LVTs) introduce local attention for reducing computational complexity and local-global interactions for performance enhancement.

The fixed-size input RGB image $IF$$\in$$\mathbb{R}^{H\times W\times 3}$ is first partitioned into non-overlapping patches of size $a\times a$, resulting in $\frac{H}{a}\times \frac{W}{a}$ visual patches with feature set as the concatenation of the original pixel RGB values.
These patches are then projected into an arbitrary dimension $C$ using a linear embedding layer.
Patch merging is applied in stage $s\in[2,S]$ to reduce the number of patches by $2 \times$ down-sampling of resolution, and the linear layer is applied to the concatenated features while setting the output dimension of $s$-th stage to $C_s=C\cdot2^{s-1}$.
As the network goes deeper, patch merging is repeatedly employed, so the resolution of $s$-th stage is $P_s=\frac{H}{a\cdot2^{s-1}} \times \frac{W}{a\cdot2^{s-1}}$.
By varying $C_s$ and $N^{bk}_s$, the model size and complexity can be scaled accordingly.
Differently, there is no patch merging and downsampling for ViT model.

\textbf{Transformer block (encoder):}
Each transformer block consists of a linear projection (LP), an MSA module, and a feed-forward network (FFN) containing two fully-connected (FC) layers.
Layernorm (LN) layer is applied before MSA and FFN, and a residual connection is applied after each module.

In LP, the input matrix is projected into three spaces, Query($Q$), Key($K$), and Value($V$), by multiplying the weights ($W^Q$, $W^K$, and $W^V$).
Moreover, different multiple groups of QKV corresponding to multiple heads of MSA are generated.
In each head, $Q$ is multiplied by $K$ to obtain the attention matrix, which represents the relevance of each two patches. Each row of attention matrix is normalized to probabilities (attention scores) using the softmax function. Finally, the scores are used to weighted sum $V$ to obtain the output embedding.
These GeMMs are parameter-free, unlike the parameterized matrix multiplications in FC layers of MSA and FFN.
Then, FC layer takes the output embedding of all attention heads as input and performs a linear projection.
Thus, for each transformer block, there are $(3 + 1 + 2\cdot 4)\cdot C_s^2=12\cdot C_s^2$ parameters, including $W^Q$, $W^K$, $W^V$, $W^{\text{MSA\_FC}}$, $W^{\text{FFN\_FC1}}$, and $W^{\text{FFN\_FC2}}$.

\textbf{Local attention:}
For LVTs, to reduce the complexity linear to the input size, all input patches of each transformer block at $s$-th stage are divided into $\frac{H}{a\cdot2^{s-1}\cdot R_h} \times \frac{W}{a\cdot2^{s-1}\cdot R_w}$ small-sized local regions of size $R_h\times R_w$, each of which independently serves as an input to the computational branch based on local attention.
As in Swin\cite{Swin}, many and uneven computational branches appear at each stage, 1225, 324, 81, and 25, respectively, as the input is 960$\times$960 and both $R_h$ and $R_h$ are 7.
Note that the computational procedures and weights are the same for all branches, and there are no data dependencies between them, except for the local-global interactions.

\begin{figure}[htbp]
\centering
\includegraphics[width=0.39\textwidth]{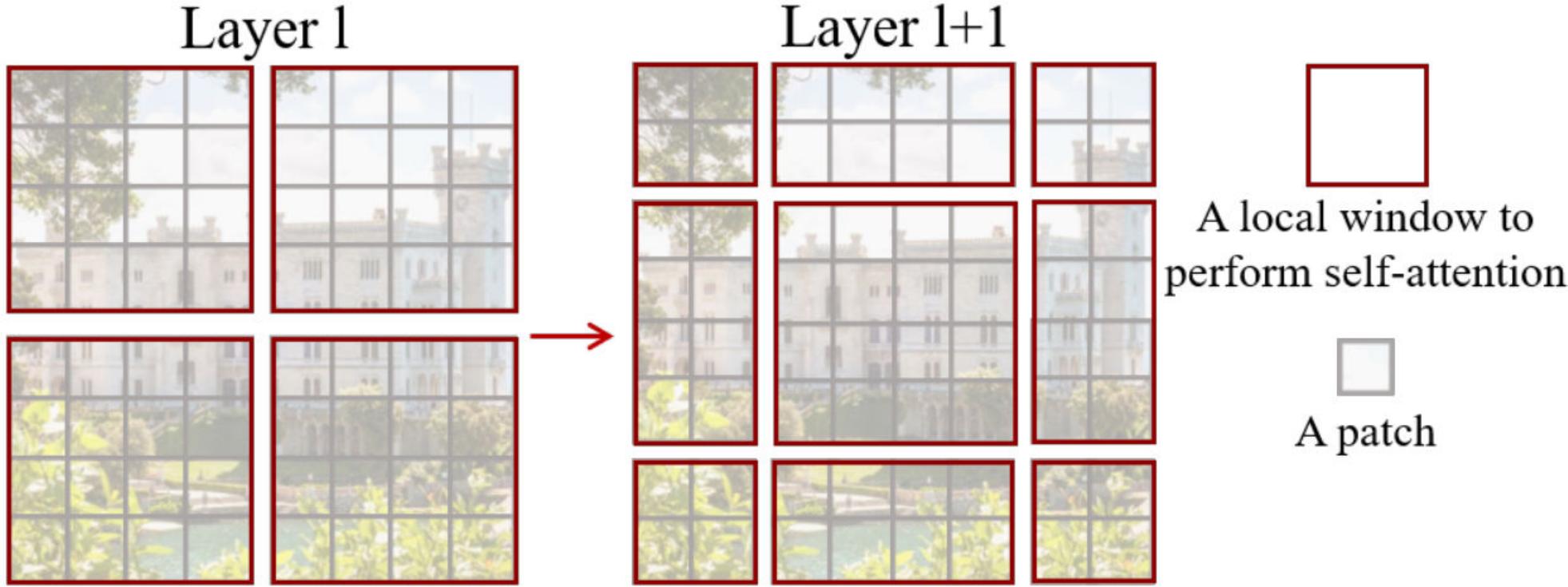}
\caption{Shifted window partitioning for local-global interaction\cite{Swin}.}
\label{fig:shift}
\end{figure}

\begin{figure}[htbp]
\centering
\includegraphics[width=0.37\textwidth]{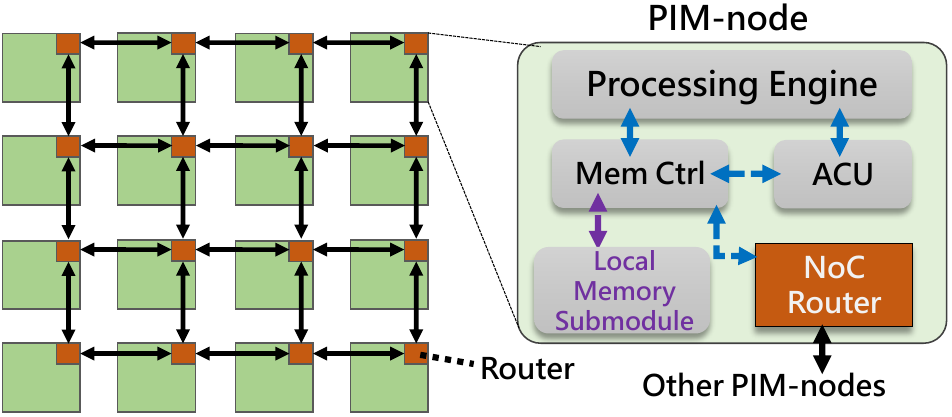}
\caption{Scalable processing in-memory systems.}
\label{fig:pim}
\end{figure}

\begin{figure}[!t]
    \centering
    \includegraphics[width=0.48\textwidth]{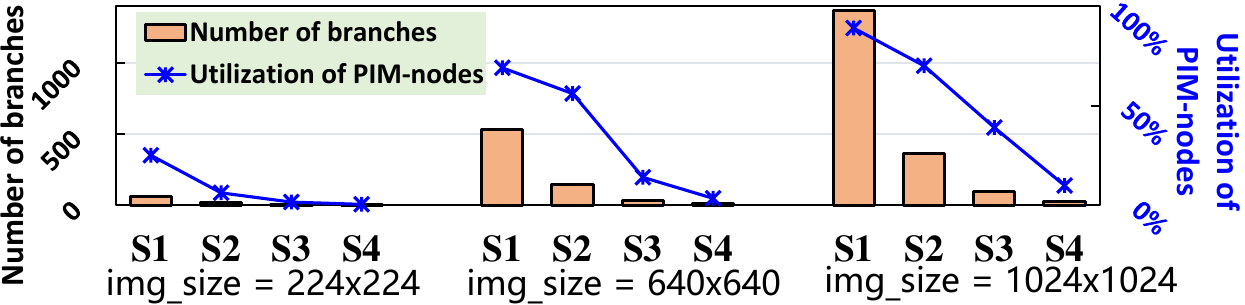}
    \caption{PIM-node utilization under branch-level parallelism. Swin is deployed to the PIM system with a node array of size 16$\times$16, when the input is $640^2$ and the local region is $7\times 7$. During the execution of the most computationally intensive stage 3 and 4, the utilization is severely below 15\%.}
    \label{fig:Branch-level}
\end{figure}

\textbf{Local-global interaction:}
\label{sec:local-global}
Before that, different computational branches extract the local information of different regions.
To regain the ability to understand the global context and long-term dependencies of the whole input, LVT models exert great efforts in implementing local-global interactions, which is to interchange feature matrices or $KV$ matrices between local regions by some sophisticated operations.
Some models adopt a dual-block architecture, with the first block using local region-based attention and the second block enabling cross-region interaction through shifted window partitioning (Swin\cite{Swin}) or global sub-sampling (Twins\cite{Twins}) or spatial shuffle operations (Shuffle\cite{shuffle}).
Besides, other models work to achieve satisfactory receptive fields within every transformer block in a parallel manner, such as complementary coarse-grained global patches (Focal\cite{Focal}) or cross-shape window (CSWin\cite{CSWin}).

\subsection{Processing In-Memory Systems}
\label{sec:PIM}

The paradigm of PIM architecture is depicted in Figure \ref{fig:pim}, which employs a spatial structure, a scalable PIM-node array connected via network-on-chip (NoC).
The spatial structure has been developed in the domain of DNN acceleration by industry and academia \cite{TANGRAM,TETRIS,Neurocube,DOJO,Simba,Atomic,inter-layer-scheduling}.
NoC is widely used as an interconnect fabric for DNN accelerators due to its good scalability and energy efficiency to handle rapidly evolving DNNs\cite{noc}.

We aim to abstract the basic hardware requirements (templates) for the PIM system commonly used in DNN domains\cite{TETRIS,Neurocube,DOJO,VLSI}, to support our research in visual Transformer deployment, as has been done in CNN deployment\cite{nicepim,DDAM}.
The description of PIM architecture is as follows:

$\bullet$ Each PIM-node with integrated compute and memory comprises a processing engine, a router, an auxiliary computing unit (ACU), a memory controller, and a memory submodule, which can be MiB-sized DRAM banks/subarrays or SRAM. All memory and compute resources are distributed across the PIM-node array.

$\bullet$ Unlike normal spatial architectures, the entire weights of the model are loaded on the on-chip distributed memory of the PIM system prior to the inference execution.

$\bullet$ Each PIM-node can access its local memory submodule directly, or access remote memory submodules of other PIM-nodes via a mesh-based NoC.

$\bullet$ Each processing engine consists of a processing element (PE) array, which can be a systolic array\cite{TPU}, NVDLA-style array (parallel vector MAC units)\cite{NVDLA}, or coarse-grained reconfigurable array (CGRA)\cite{BusMap}, etc, along with SRAM buffers for inputs, weights, and outputs to do multiply-and-accumulate (MAC) operations.
ACU serves to do nonlinear computations, such as softmax, layernorm, and gelu.

Specifically, the scalable PIM system has a PIM-node array shaped as $H_A\times W_A$, each PIM-node has a PE array of $r\times r$, each memory submodule with a capacity of $node\_cap$ and bandwidth of $bw$, and the NoC with a link bandwidth of $BW$.

\subsection{Challenges with End-to-end Inference Deployment}
\label{sec:challenge}
\textbf{1. How to parallelize many computational branches:}
As the trendiest visual Transformers, LVTs have many computational branches based on local attention in each transformer block.
However, branch (local region)-level parallelism, where each computational branch is processed on only one PIM-node, results in underutilization of PIM-nodes due to the coarse-grained partitioning, as shown in Figure \ref{fig:Branch-level}.

\textbf{2. How to fully exploit limited and distributed on-chip memory resources:}
Most models present a hierarchical structure, and transformer blocks in different stages have distinct computational branches and workloads.
The entire parameters of a model need to be stored on on-chip limited memory.
Storing the parameters duplicated across multiple PIM-nodes reduces data movement and improves data locality, but it causes tighter on-chip memory resources.
For models with tens of millions of parameters or more, the memory submodule on PIM-node is MiB-sized and cannot store the whole parameters.
Besides, each PIM-node has to allocate enough workspace for intermediate results so that the computational branches on each transformer block have to be batch-processed.

\textbf{3. How to fulfill low-cost local-global interactions:}\label{sec:problem3}
For local-global interactions, different branches exchange their local feature matrices or KV matrices in a fixed pattern.
During deployment, one should optimize the arrangement of computational branches to minimize the distance of branches with dependencies for low-cost information interactions.

\section{Framework Overview}
\label{sec:overview}

\begin{figure}[tbhp]
\centering
\includegraphics[width=0.47\textwidth]{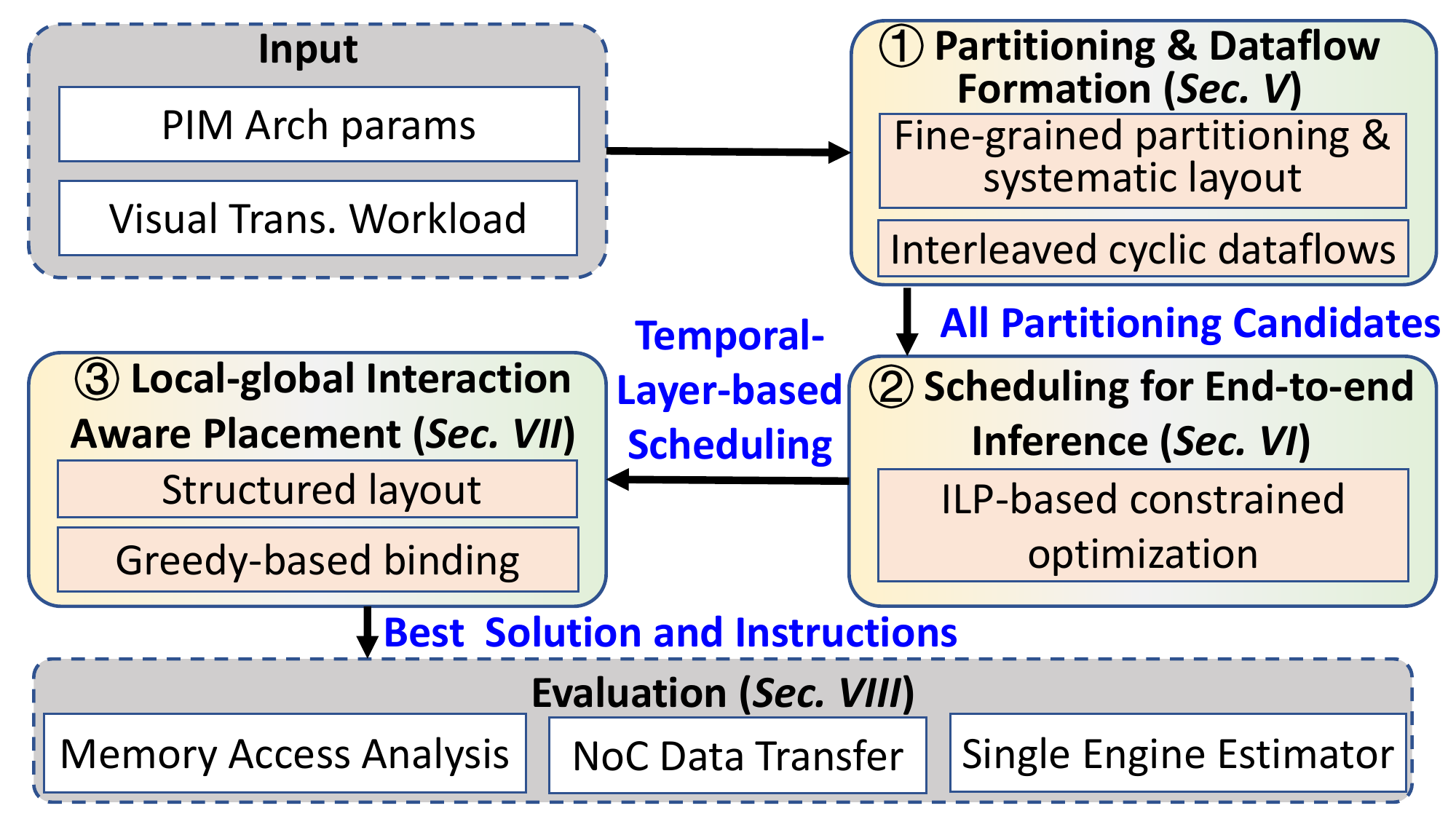}
\caption{Allspark overview.}
\label{fig:overview}
\end{figure}

The deployment framework, Allspark, is proposed to fulfill end-to-end inference with minimum latency for visual Transformer models on PIM systems. The overview is shown in Figure \ref{fig:overview}, which mainly consists of three parts: branch-oriented partitioning and dataflow formation, scheduling for end-to-end inference, and local-global interaction aware placement.
Given a visual Transformer model and a detailed configuration of PIM architecture, Allspark fully automatically generates the optimal deployment scheme and corresponding hardware instructions for evaluating the cost of memory accesses, NoC data transfers, and intra-node processing.
Since the input image is fixed-sized, which means the workload is deterministic, Allspark generates the deployment solution at compile-time.

\textbf{Partitioning}. To fulfill massively parallelism on the PIM system, Allspark employs a fine-grained partitioning scheme for computational branches, and format a systematic layout and NoC-based interleaved dataflow, which realizes operator fusion-like effects on each PIM-node.
That is, the output of the preceding operator is stored in the local memory submodule and could be fetched/reused as input by the next operator, reducing frequent data movement between PIM-nodes and improving data locality.
The proposed scheme would provide all partitioning candidates for all computational branches.

\textbf{Scheduling}. Optimal exploitation of distributed computing and memory resources is the primary concern. To run the full model on a resource-limited distributed PIM system and minimize the inference latency, Allspark formulates static scheduling as a constrained optimization problem. The optimal solution is selected from all candidate partitions, and all computational branches are temporally and spatially scheduled.

\textbf{Placement}. To cope with complex dependencies for local-global interactions, Allspark provides a detailed placement of computational branches, after scheduling a whole model.
The reason is that the interaction communication cost is quite small relative to within-branch computation cost under ideal congestion-free NoC traffic conditions.
For instance, the complexity of a local region is about $12nC^2 + 2n^2C$, and its local feature map is of size $nC$.
Assuming a PE array size of $8\times8$ and a NoC width of 64 bits, a rough estimate indicates that, under ideal conditions, the time required for data interchange in local-global interactions is within 1\% of all computation time.
Hence, for detailed placement, each computation branch is assigned to a PIM-node sub-array while minimizing the communication cost under global-local interactions.

\section{Partitioning and Dataflow Formation}
\label{sec:dataflow}
\begin{figure*}[!t]
\centering
\includegraphics[width=0.8\textwidth]{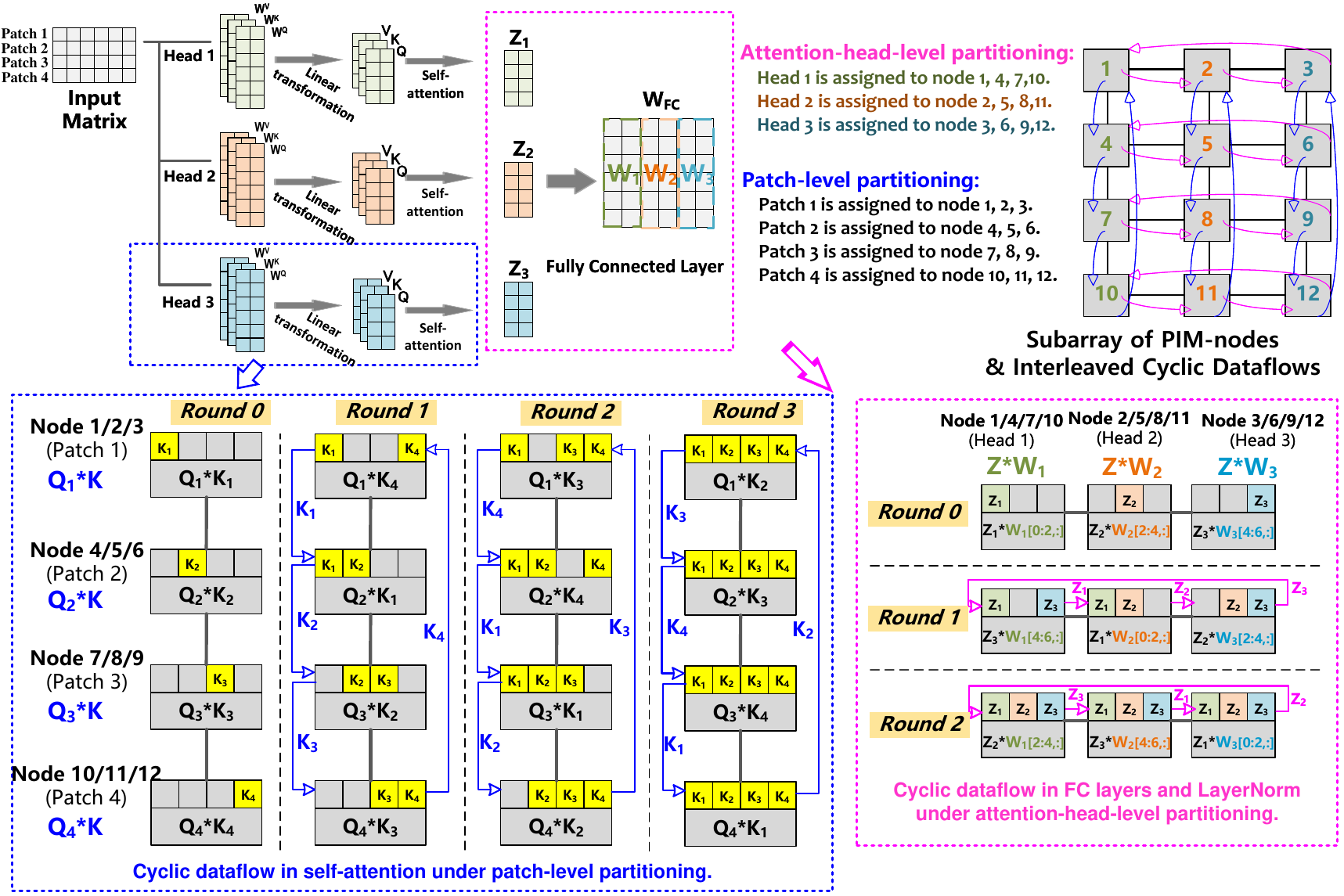}
\caption{Attention-head-level partitioning and patch-level partitioning (Take $N=4$, $h=3$, $u=4$, and $v=3$ as an example).}
\label{fig:parallel}
\end{figure*}

To parallelize each computational branch based on local attention, by virtue of the flexibility of NoC, we propose a flexible and fine-grained partitioning, a systematic layout, and interleaved cyclic dataflows (see Figure \ref{fig:parallel}).
This enables an operator fusion-like effect in the case of massive parallelism, reducing data movement to improve data localization.

\subsection{Key Ideas}
\label{text:partition1}

\textbf{Fine-grained partitioning and systematic layout:}
All the computational branches are isolated from each other as inputs to an encoder, which has two distinct properties, firstly, the inputs and outputs at different encoder layers are related to patches within the window, and secondly, the computation of all attention heads within MSA is completely independent.
To reuse data and reduce data movement, we propose a uniform and flexible partitioning for all encoder layers along with \emph{attention-head granularity} and \emph{patch granularity}. That is, the matrix multiplications on each layer are uniformly sliced into smaller chunks and then assigned to a PIM-node subarray of \emph{variable} size $u \times v$ for parallel processing, and $u, v \geq 1$.

Under attention-head-level partitioning, we assign all $h$$(=C/d)$ attention heads of MSA to $v$ groups of nodes, each group being lined up by $u$ nodes and being given $b$ ($=\lceil h/v\rceil$, $h\geq v$) heads.
In each attention head, the inputs and outputs (e.g., feature matrices, QKV) of all layers consist of patch-related vectors, and there are $N(=R_h \times R_w)$ patches in each local region.
Under patch-level partitioning, these patch-related vectors are equally allocated across $u$ PIM-nodes, each dealing with $p$ vectors.
Finally, feature matrix $\mathbf{F}\in\mathbb{R}^{N\times C}$ is partitioned and each node holds a fraction $\mathbf{F}'\in\mathbb{R}^{p\times q}$, where $p=\lceil N/u\rceil$ and $q=b\cdot d=C/v$.

\textbf{Interleaved cyclic dataflows:}
The inter-layer cyclic dataflow is used for broadcasting different parts of data/weight when a single DNN layer is distributed over multiple nodes\cite{TANGRAM}. Differently, in this work, we devise interleaved cyclic dataflows for both MSA and FC layers based on consistent partitioning and systematic layout.
MSA requires each PIM-node to receive partial $K$ and $V$ matrices generated from the remaining $u-1$ nodes, and broadcast its own partial matrix to them.
For the cyclic data transmission scheme, each node only transmits the partial $K\&V$ matrices to the nearest succeeding node, except for the last node, which exclusively transmits it to the first node, forming a cyclic path.
Additionally, for FC layers, the embedding matrices generated by different attention heads are concatenated and multiplied by the weight matrix.
Since each embedding vector is scattered over $v$ nodes in different groups, we also employ a cyclic data path to propagate embedding matrices for GeMM parallelism.

Specifically, on each PIM-node, the output of the preceding operator is directly consumed by the next one, which increases data reuse and reduces data movement.
Subtly, after each process phase based on the cyclic dataflows, the output still maintains the same distribution as the input, and each PIM-node always holds a partial size $p\times q$.
Thanks to the tailored systematic layout, these interleaved cyclic dataflows stream for data multicasting form a pattern of alternating \textit{vertical} and \textit{horizontal} circulation (top right of Figure \ref{fig:parallel}), which avoids contention for NoC bandwidth and relieves traffic pressure.

\vspace{-0.1cm}
\subsection{NoC-based Dataflow Implementation}
\textbf{Linear transformation:}
Initially, each PIM-node holds a partial feature matrix $\mathbf{F}'\in\mathbb{R}^{p\times q}$, and can further access the $\mathbf{F}_{p\times C}$ through $v-1$ rounds of \textit{horizontal} cyclic dataflows.
For each attention head, the weights $W^Q$, $W^K$, and $W^V$$\in\mathbb{R}^{C\times d}$ are pre-stored on each counterpart PIM-node.
On each node, $\mathbf{F}_{p\times C}$ is multiplied by these weight matrices to obtain three sub-matrices $Q'$, $K'$, and $V'\in\mathbb{R}^{p\times d}$, respectively.
Then, $Q$, $K$, and $V\in\mathbb{R}^{N\times d}$ are distributed across $u$ PIM-nodes.

\textbf{Attention:}
For each attention head, $softmax(Q_{p\times d}\cdot K^T_{N\times d})\cdot V_{N\times d}$ is calculated on each node.
Although only $K'_{p\times d}$ and $V'_{p\times d}$ are stored on each node, the fully-size $K$ and $V$ distributed over all $u$ nodes are available to each node, using $u-1$ rounds of \textit{vertical} cyclic dataflows (bottom left of Figure \ref{fig:parallel}).
In each round, each node sends only a partial matrix of size $p\times d$ to the succeeding node.
Differently, each node sends one locally generated in the first round, and sends the one received from the previous round in subsequent rounds.

The assigned $b$ attention heads are processed sequentially on each node, and the above steps are repeated $b$ times to obtain the output matrix $Z_{p\times q}$.
Ultimately, the volume of the partial $\mathbf{F}$ or $K$ or $V$ matrix to be transmitted at each node is $D^{\mathbf{F}}=D^{K} = D^{V} =  b \times (u-1) \times p \times d=  (u-1) \times p \times q$.

\label{sec:FC}
\textbf{Fully-connected layer and patch merging:}
After the self-attention, each node starts with only a partial matrix $Z_{p\times q}$, and we use $v-1$ rounds of \textit{horizontal} cyclic dataflows so that each node acquires the matrices $Z_{p\times aC}$ distributed across $v$ nodes, as shown in the bottom right of Figure \ref{fig:parallel}.
For FC layer and patch merging, each node calculates $Z_{p\times aC}\cdot W'_{aC\times q}=O_{p\times q}$, where $a\geq1$ expresses the scaling of the weight size.
The full-size weight matrix $W_{C\times aC}$ or $W_{aC\times C}$ is also pre-partitioned and pre-stored to the counterpart $v$ node groups, respectively, and all $u$ nodes in the same group store the same $W'_{aC\times q}$.
Note that no weight matrices and intermediate results are transferred between nodes in these phases.
After computation, the full-size output matrix $O$ remains evenly distributed across the node subarray of size $u\times v$, so subsequent FC layers can still be executed using the cyclic data transfer described above.

\textbf{LayerNorm:}
LayerNorm is to normalize the embedding vector of each patch, though each vector is sliced into $v$ PIM-nodes, and each node's ACU calculates the mean and standard-deviation of the vector segment it holds.
These two intermediate result matrices of size $p \times 1$ are aggregated from the other nodes to each node during the computation, using $v-1$ rounds of \textit{horizontal} cyclic dataflows in Section \ref{sec:FC}.

Finally, for one FC layer of MSA, two FC layers of FFN, layernorm, and patch merging, the amount of data transmitted per PIM-node is $D^{\text{MSA\_FC}} =  (v-1) \times p \times q$, $D^{\text{FFN}} = (v-1) \times p \times q +(v-1) \times p \times a_1q$, $D^{LN} = 2\times (v-1) \times p \times 1$, and $D^{\text{PM}}=(v-1) \times p \times a_2q $, where $a_1=4$ and $a_2=2$.

\vspace{-0.5em}
\subsection{Processing Procedures and Weight Burdens per PIM-node}
\begin{figure}[!t]
\centering
\includegraphics[width=0.5\textwidth]{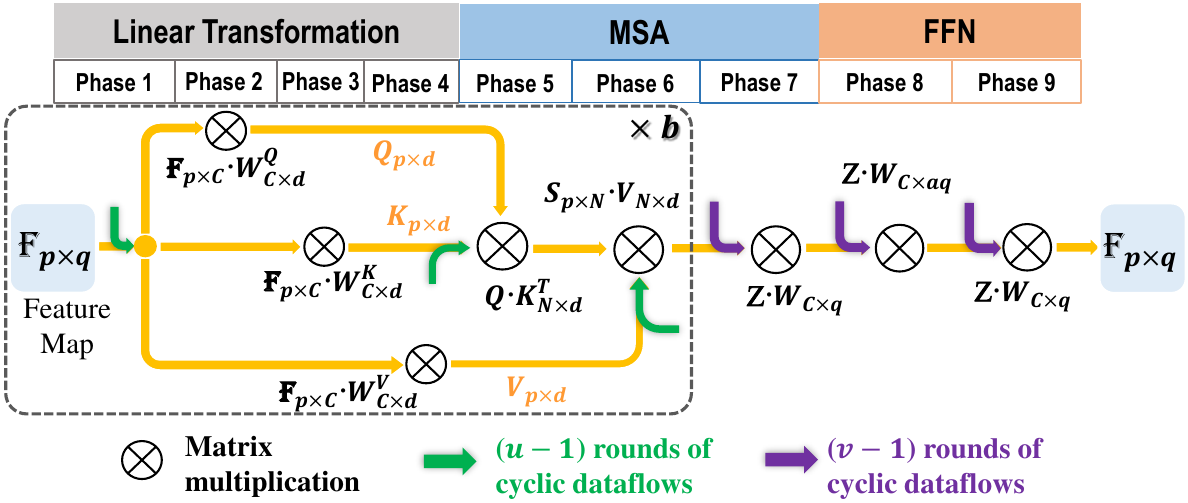}
\caption{Processing procedures on each PIM-node.}
\label{fig:single-node}
\end{figure}

Each PIM-node of the subarray executes the same computational procedure consisting of nine phases, as shown in Figure \ref{fig:single-node}.
At each phase, the PE array on every PIM-node performs a specified matrix multiplication.
Then, in some phases, each PIM-node needs to retrieve partial matrices scattered across others, utilizing the cyclic dataflows described above.

Briefly, all the weights used by each branch (totaling $12\cdot C_s^2$) are stored in the PIM sub-array of size $u\times v$.
Once a individual branch is partitioned into $v$ groups of nodes on attention-head level, each group of nodes will use distinct weights, so we tentatively divide each of the weighting matrices into $v$ parts, with only one part pre-stored for each group of PIM-nodes.

\section{Scheduling for End-to-end Inference}
\label{sec:partition}
\begin{figure}[!t]
\centering
\includegraphics[width=0.4\textwidth]{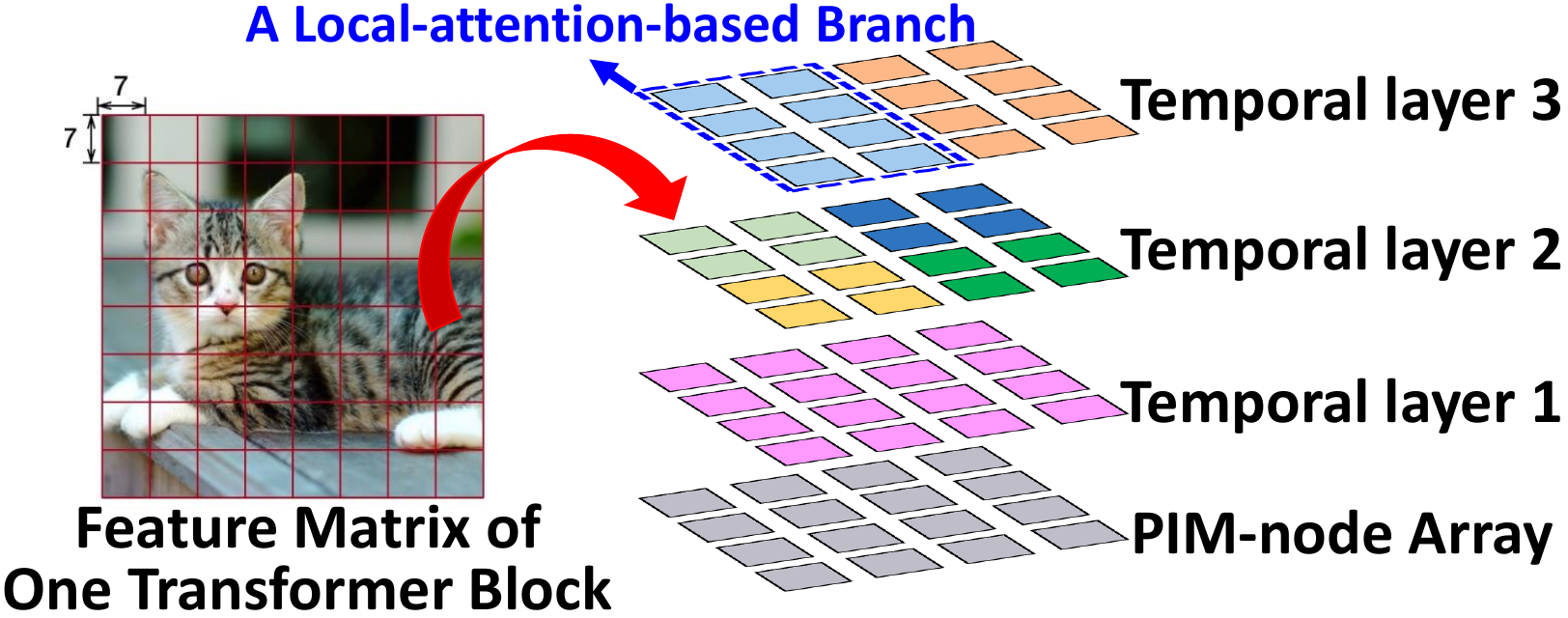}
\caption{Scheduling of each encoder.}
\label{fig:scheduling}
\end{figure}

To accelerate model inference, pipeline parallelism and tensor parallelism are widely used strategies\cite{FlexGen}.
Typically, the former yields high throughput and the latter enables low inference latency.
To minimize the inference latency, we propose a scheduling method based on tensor parallelization, which sequentially processes all encoders of the whole model.
Besides, each encoder is partitioned to occupy all distributed resources on the PIM.
Accordingly, given partitioning candidates of all computational branches, the proposed scheduler assigns a complete model temporally and spatially to the PIM system.
The main strategies in Figure \ref{fig:scheduling} are as follows:

\textbf{1) Consistent partitioning:} Since encoders within the same model stage have identical amounts of computational branches and workloads, consistent partitioning and scheduling are applied to them. Moreover, all are executed sequentially, with each one utilizing all PIM-nodes via tensor parallelism.

\textbf{2) Sub-lot processing for computational branches:} Each branch can be parallelized on a PIM-node subarray, but the system probably cannot handle all branches on an encoder concurrently. Besides, combining the partitioning candidates of all computational branches results in a huge solution space. To simplify the matter, for all branches with consistent computation, we execute all of them in multiple lots, with each lot occupying all PIM-nodes at the same timespan.

\textbf{3) Equal partitioning of branches in each lot:}
In the case of different partitioning schemes for computational branches in the same lot, there occurs unbalanced workloads and idle computational resources.
Thus, we employ an equal partitioning scheme for branches within the same lot, and we define the parallelization of each timespan as a \emph{\textbf{temporal layer}}.
There are multiple temporal layers as all branches are organized into several lots.
For more flexible scheduling, different partitioning schemes can be used between temporal layers.
For example, branches in two temporal layers could be assigned to node subarrays of size $2\times 2$ or $3\times 4$, respectively.

\textbf{4) Memory capacity constraint:} For each PIM-node, the associated memory submodule shall store all weight matrices used and the intermediate results generated during inference execution.
Therefore, the memory space required MUST NOT exceed the capacity of the memory submodule per PIM-node.

\begin{table}[tbhp]
  \centering
  \caption{Key notation.}
  \label{table:notation}
  \resizebox{90mm}{11mm}{
  \begin{tabular}{cc|cc|cc}
    \toprule
    \multicolumn{2}{c|}{\textbf{Variables}} & \multicolumn{2}{c|}{\textbf{Constants}}& \multicolumn{2}{c}{\textbf{Indices}} \\
    \midrule
    \textbf{X}& binary matrix to& $N^{tl}$& temporal layers & $s$ & model stage\\
     & represent a schedule&  $N^{br}$&  branch counts & $i$ & branch index\\
    & & $N^{bk}$& block counts& $j$ & temporal layer\\
    \textbf{Y}& auxiliary binary matrix& $u$& subarray rows & $\alpha$  & $U_{s}$ index\\
    \textbf{Z}& auxiliary binary matrix& $v$& subarray columns & $\beta$  & $V_{s}$ index\\
    \bottomrule
  \end{tabular}
  }
\end{table}

Furthermore, we formulate the problem as an integer linear programming (ILP).
This allows us to determine the number of temporal layers for each encoder, and the number of computational branches and the partitioning schemes on each temporal layer.
In addition, to alleviate the memory burden of each PIM-node, we propose weight sharing and weight reuse.

\subsection{Variables and Constants}
Binary variable $X_{s,i,j,\alpha,\beta}$ indicates whether there are a total of $i$ branches in $j$-th temporal layer of each Encoder in $s$-th model stage, and each branch is assigned to a PIM-node sub-array of variable size $u_{s,j}\times v_{s,j}$, where $i\leq N_s^{br}$, $j\leq N_s^{tl}$, $s\leq S$.
$N_s^{tl}$ and $N_s^{br}$ mean temporal layer counts and branch counts on each encoder in $s$-th stage, respectively.
As every branch entirely occupies a temporal layer, this implies the most amount of temporal layers, so $N_s^{tl}=N_s^{br}$.

Following the partitioning candidates for all computational branches, the computational branch containing $\mathrm{H}$ attention heads of MSA and $\mathrm{P}$ patches is assigned to a sub-array of variable size $u_{s,j}\times v_{s,j}$.
Furthermore, $u_{s,j}$ takes from $\{1,2, ..., \lceil \frac{\mathrm{H}}{2}\rceil, \mathrm{H}\}$ as an array $U_{s}$, and $v_{s,j}$ takes from $\{1,2, ..., \lceil \frac{\mathcal{P}}{2}\rceil, \mathcal{P}\}$ as another array $V_{s}$, where $\mathcal{P}=\frac{\mathrm{P}}{p_{min}}$ and $p_{min}$ is the minimum granularity of patch chunks (for the PE array size).
Here, $\alpha$ and $\beta$ are the indexes of $U_{s}$ and $V_{s}$ respectively, namely $u_{s,j}=U_{s}[\alpha]$ and $v_{s,j}=V_{s}[\beta]$.
Integer variables $V^{dyn}_{s,j}$, and auxiliary binary variables $Y_{s,j,j',\alpha,\beta}$ and $Z_{s,j,\alpha,\beta}$ in the below subsection are presented.

\vspace{-1em}
\subsection{Constraints}
These variables and constants serve to express constraints derived from the above scheduling strategies:
\begin{align}
\centering
    \label{eq:cons-1}
    \enspace &\sum\nolimits_{i}{X_{s,i,j,\alpha,\beta}}\leq N_{\alpha,\beta}\\
    \label{eq:cons-2}
    &\sum\nolimits_{i,j,\alpha,\beta}{X_{s,i,j,\alpha,\beta}}\cdot i= N^{br}_s\\
    \label{eq:cons-3}
    &\sum\nolimits_{i,\alpha,\beta}{X_{s,i,j,\alpha,\beta}}= 1\\
    \label{eq:cons-4}
    &B_{s,j}\geq B_{s,j+1} \ \ \text{and} \ \ B_{s,j}=\sum\nolimits_{i,\alpha,\beta}{X_{s,i,j,\alpha,\beta}}\cdot i \\
    \label{eq:cons-5}
    &\sum\nolimits_{s}{V^{wt}_{s}\cdot {N^{bk}_s}} + \mathop{max}\limits_{s,j}{\{V^{dyn}_{s,j}\}} \leq node\_cap
\end{align}

Constraint \ref{eq:cons-1} is that branches on $j$-th temporal layer are assigned to the node subarrays of size $u_{s,j}\times v_{s,j}$, respectively, and the number of branches assigned cannot exceed the upper limit $N_{\alpha,\beta}$, which is the maximum number of subarrays with size $u_{s,j}\times v_{s,j}$ that can be packed by the PIM system.
Section \ref{sec:layout} states how to derive the value of $N_{\alpha,\beta}$.
Constraint \ref{eq:cons-2} is that all branches of each encoder must be assigned, and that each branch can only be assigned to one temporal layer.
Constraint \ref{eq:cons-3} is that all branches on each temporal layer are identically partitioned and assigned to a node sub-array with the same size.
Constraint \ref{eq:cons-4} is that the number of branches allocated to different temporal layers decreases progressively.

Constraint \ref{eq:cons-5} implies that, under the current scheduling solution, the sum of the stored weights and the required workspace should not exceed the memory submodule's capacity on each PIM-node.
The required workspace is used to store the input/output feature maps and intermediate data during the computation of each computational branch.
For each PIM-node, $V^{wt}_{s}$ is the volume of weight parameters that need to be stored for each block in the $s$-th model stage, while $V^{dyn}_{s,j}$ is the volume of workspace required during the computation of the $j$-th temporal layer in the $s$-th model stage.
According to Figure \ref{fig:single-node}, for each temporal layer, the computation per node consists of nine phases.
Notably, phases 2, 6, and 8 exhibit a higher demand for workspace, surpassing the other phases.
Thus, $V^{dyn}_{s,j}$ = $\mathop{max}\limits_{p=1,...,9}\{V^{dyn}_{s,j,p}\}$$\in$$[0,node\_cap]$, which is easily rewritten as linear expressions with some auxiliary variables.

However, if we were to pre-store all weights associated with the computations assigned to each node across all temporal layers, it would impose a substantial storage burden.
For PIM systems with limited memory resources, constraint \ref{eq:cons-5} is so harsh that either no feasible solution exists or the solution is of poor quality with high inference latency.

\subsection{Memory Constraint-driven Weight Sharing and Reuse}
\label{text:weight_reuse_sharing}
To alleviate the memory burden on each PIM-node, we propose weight sharing and weight reuse.
The former refers to the sharing of weights across nodes on a sub-array on the \textit{same temporal layer}, whereas the latter relates to the reuse of weights among nodes across \textit{different temporal layers}.

\textbf{Weight sharing:}
At each temporal layer, each branch is assigned to a PIM-node subarray of size $u_{s,j}\times v_{s,j}$, on which the weight parameters of $W_{C_s\times q}^{\text{PM}}$, $W_{C_s\times d}^Q$, $W_{C_s\times d}^K$, $W_{C_s\times d}^V$, $W_{C_s\times q}^{\text{FC}}$, and $W_{C_s\times aq + C_s\times q}^{\text{FFN}}$ need to be stored.
Specifically, as shown in Figures \ref{fig:parallel} and \ref{fig:single-node}, in attention-head-level partitioning, the weight parameters are equally distributed among the $v_{s,j}$ node sets.
Moreover, in patch-level partitioning, each set consisting of $u_{s,j}$ PIM-nodes stores identical weight parameters.

We distribute these identical weights equally across the $u_{s,j}$ PIM-nodes.
Then, $u_{s,j}-1$ rounds of cyclic dataflows are executed before the phase 2$\sim$4 and 7$\sim$9 of each computation, allowing each node to share its own stored weights to other nodes.
To minimize the cost of weight sharing, the weights at each phase are used for batch inputs and released promptly after each phase to minimize memory usage.

Ultimately, the volume of weights to be transmitted at each node are $W^{{\text{PM}}}_{\alpha,\beta}$,  $W^{{\text{MSA}}}_{\alpha,\beta}$, $W^{{\text{FC}}}_{\alpha,\beta}$, and $W^{{\text{FFN}}}_{\alpha,\beta}$, respectively.

\textbf{Weight reuse:}
All branches of the same encoder use the same weights, but each node stores only a partial one, the below cases allow to reuse weights across temporal layers:
\begin{enumerate}
  \item If, at a specific temporal layer, the branch is assigned to only one PIM-node, i.e., $u_{s,j}=v_{s,j}=1$, then this PIM-node stores all the weight parameters;
  \label{reuse:cd-1}
  \item If branches on two temporal layers of the same Encoder follow the same partitioning scheme.
  \label{reuse:cd-2}
\end{enumerate}
These cases can be expressed as follows:
\begin{align}
\centering
\fontsize{10}{10}\selectfont
V^{wt}_{s} = \begin{cases}
12\cdot C_s^2, \quad\ \exists\ i,j;\,X_{s,i,j,0,0}=1\; (\textbf{Term} \ref{reuse:cd-1})\\
\sum_{j}{vol_{s,j}}, \quad\text{otherwise}.
\end{cases}
\end{align}
\vspace{-1em}
\begin{align}
\centering
\fontsize{10}{10}\selectfont
vol_{s,j} = \begin{cases}
0, \quad\quad \exists\ j'\ \text{and}\ 0<j'<j<N^{tl}_s,\\
\quad\quad\quad \sum_{i}{X_{s,i,j,\alpha,\beta}} = \sum_{i}{X_{s,i,j',\alpha,\beta}};\\
\quad\quad\quad\quad\quad\quad\quad\quad\quad\quad\quad(\textbf{Term}\ \ref{reuse:cd-2}) \\
\frac{12\cdot C_s^2}{U_{s}[\alpha]\times V_{s}[\beta]},\quad\quad \text{otherwise}.
\end{cases}
\end{align}

By introducing two auxiliary binary variables $Y_{s,j,j',\alpha,\beta}$ and $Z_{s,j,\alpha,\beta}$, the above can be transformed into a linear expression:
\begin{align}
\raggedleft
\fontsize{10}{10}\selectfont
\begin{cases}
& 1-\sum_i X_{s, i, j^{\prime}, 0,0} \geq Y_{s, j, j^{\prime}, \alpha,\beta}; \\
& \sum_i X_{s, i, j, \alpha,\beta}-\sum_i X_{s, i, j^{\prime}, \alpha,\beta}-\sum_i X_{s, i, j^{\prime}, 0,0}\\
    &\quad\quad\quad\quad\quad\quad\quad\quad\quad\quad\quad\quad\quad\quad  \leq Y_{s, j, j^{\prime}, \alpha,\beta}; \\
& \sum_i X_{s, i, j, \alpha,\beta} \geq Y_{s, j, j^{\prime}, \alpha,\beta}; \\
& 1-\sum_i X_{s, i, j^{\prime}, \alpha,\beta} \geq Y_{s, j, j^{\prime}, \alpha,\beta};\\
& 0 \leq j^{\prime}<j<N^{tl}_s.
\end{cases}
\end{align}
\vspace{-0.8em}
\begin{align}
\raggedleft
\fontsize{10}{10}\selectfont
\begin{cases}
& j-\sum_{j^{\prime}=0}^{j-1} Y_{s, j, j^{\prime}, \alpha,\beta} \leq j \cdot Z_{s, j, \alpha,\beta}; \\
& \sum_{j^{\prime}=0}^{j-1} Y_{s, j, j^{\prime}, \alpha,\beta}+Z_{s, j, \alpha,\beta} \leq j.
\end{cases}
\end{align}
where $Z_{s,j,\alpha,\beta}=1$ indicates that branches at $j$-th temporal layer will reuse weights from other layers for the encoder of $s$-th model stage, and 0 otherwise.
$Y_{s,j,j',\alpha,\beta}=1$ states that the branches at $j$-th temporal layer will reuse weights from $j'$-th temporal layer (meeting case \ref{reuse:cd-1} or \ref{reuse:cd-2}); and 0 otherwise.

With reuse, weights for each encoder stored on each PIM-node in the $s$-th stage is expressed as:
\begin{align}
\centering
&V^{wt}_{s} = \sum\nolimits_{j}{vol_{s,j}}\\
&vol_{s,j} = \sum\nolimits_{\alpha,\beta}{(1-Z_{s,j,\alpha,\beta})\cdot \frac{12\cdot C_{s}^2}{U_{s}[\alpha]\times V_{s}[\beta]}}
\end{align}

\subsection{Objectives}
\label{sec:obj}
All encoders shall be executed in temporal layers order, and the inference latency includes data transfer and weight sharing between nodes, and intra-node computation across all temporal layers of an entire model.
$N^{bt}$ defaults to 1, otherwise, to mitigate the latency associated with weight sharing, the weights for each processing phase on each PIM-node are used for small-batch inference. It is expressed as:
\begin{equation}
\begin{split}
\textbf{\emph{Minimize}}\ &T_{total} = \sum\nolimits_{s,j}[N^{bt} \cdot (T^c_{s,j} + T^t_{s,j}) + T^{\text{ws}}_{s,j}]
\end{split}
\end{equation}

Thanks to the well-organized systematic layout and cyclic dataflows as in Section \ref{sec:dataflow}, there is no contention for NoC bandwidth resources between the packets, and it is practical to conduct a runtime assessment at each phase based on the amount of data transferred between every two PIM-nodes.

\textbf{Weight sharing:}
As stated in Section \ref{text:weight_reuse_sharing}, prior to the weight-inclusive matrix multiplication, each node must fetch the required weights using NoC-based cyclic dataflows. The time cost of weight sharing is as follows:
\vspace{-0.4em}
\begin{align}
\begin{split}
\centering
T^{ws}_{s,j} = & \sum_{i,\alpha,\beta}{X_{s,i,j,\alpha,\beta}\cdot
                [W^{{\text{PM}}}_{s,j,\alpha,\beta} + (W^{{\text{MSA}}}_{s,j,\alpha,\beta}+}\\
                &{W^{{\text{FC}}}_{s,j,\alpha,\beta}+W^{{\text{FFN}}}_{s,j,\alpha,\beta}) \cdot N^{bk}_s]/BW.}
\end{split}
\end{align}

\textbf{Data Transfer:}
Pursuant to Section \ref{text:partition1}, the parallelization of computational branches causes feature maps to be transferred between nodes during intervals of matrix computation operations via vertical or horizontal cyclic dataflows.
\vspace{-0.3em}
\begin{equation}
\begin{split}
\centering
&T^t_{s,j} =  \sum_{i,\alpha,\beta}{X_{s,i,j,\alpha,\beta}\cdot [D^{\text{PM}}_{s,j,\alpha,\beta} + (D^{K}_{s,j,\alpha,\beta}+D^{V}_{s,j,\alpha,\beta}}\\
                &\quad\quad{+D^{\text{MSA\_FC}}_{s,j,\alpha,\beta}+D^{\text{FFN}}_{s,j,\alpha,\beta}+2\cdot D^{\text{LN}}_{s,j,\alpha,\beta}) \cdot N^{bk}_s]/BW}
\end{split}
\end{equation}

\textbf{Intra-node Computation:}
For each temporal layer, the processing engine on each node performs nine phases of matrix multiplication sequentially (as depicted in Figure \ref{fig:single-node}), and we can derive the overall computing time for each temporal layer.
\vspace{-1em}
\begin{align}
\begin{split}
\centering
T^c_{s,j} = &\sum_{i,\alpha,\beta}{X_{s,i,j,\alpha,\beta}\cdot [t^{\text{PM}}_{s,j,\alpha,\beta}+(t^{\text{LT}}_{s,j,\alpha,\beta}+t^{\text{MSA}}_{s,j,\alpha,\beta}+} \\
&{t^{\text{FFN}}_{s,j,\alpha,\beta}+t^{\text{NC}}_{s,j,\alpha,\beta}) \cdot N^{bk}_s]}
\end{split}
\end{align}

Given matrix multiplications and their dimensions for each phase, we can employ the existing search-based intra-node mapping strategies\cite{Timeloop,SCALE-Sim} for loop tiling, ordering, and spatial mapping to assess the computing time on processing engine.
GeMM expressed as three nested loops has low data reuse relative to convolutional operations, and PIM-node has a concise memory hierarchy, which allows the search to be done in a short time.
Based on the quantization methods\cite{ibert,FQ-ViT}, the delays $t^{\text{NC}}_{s,j,\alpha,\beta}$ is obtained for nonlinear computations on the ACUs.

There are totally $\sum_{s}{|U_s|\cdot |V_s|\cdot N_s^{br}\cdot(2N_s^{br}+ 1)}$ binary variables and $\sum_{s}{N_s^{br}}$ integer variables, where all variables relate to model size, and $|V_s|$ and $N_s^{br}$ rely on the input size.
We use \texttt{Gurobi}\cite{gurobi} to solve the ILP problem.
To combat possible oversized models, the complexity gets reduced by coarsening the partition granularity and shrinking $|U_s|$.

\section{Local-Global Interaction Aware Placement}
\label{sec:interaction}
Given the temporal-layer-based partitioning and scheduling of entire model, which specify the number of temporal layers $n_s^{tl}$ for each encoder, the branch count $n^{br}_{s,j}$ for each temporal layer $j=1,...,n_s^{tl}$ , and their respective partitioning schemes $[u_{s,j}, v_{s,j}]$, this section is to place computational branches in each encoder on PIM system for minimizing NoC communication cost in local-global interactions.

Differing from the general mapping problem\cite{HPCC}, here we have to determine to which temporal layer each computational branch is placed and even how these rectangular subarrays are laid out.
In addition, cutting-edge DNN mapping methods\cite{Atomic,TANGRAM} either employ a \emph{zig-zag} layout that cannot fit into our fine-grained partitioning scheme with the demand for rectangular regions, or even their exhaustive searches with high complexity cannot withstand architectures with massive nodes.
Allspark offers a two-stage method, which decomposes this challenging problem into structured layout and greedy-based binding.

\begin{figure}[!t]
\centering
\includegraphics[width=0.45\textwidth]{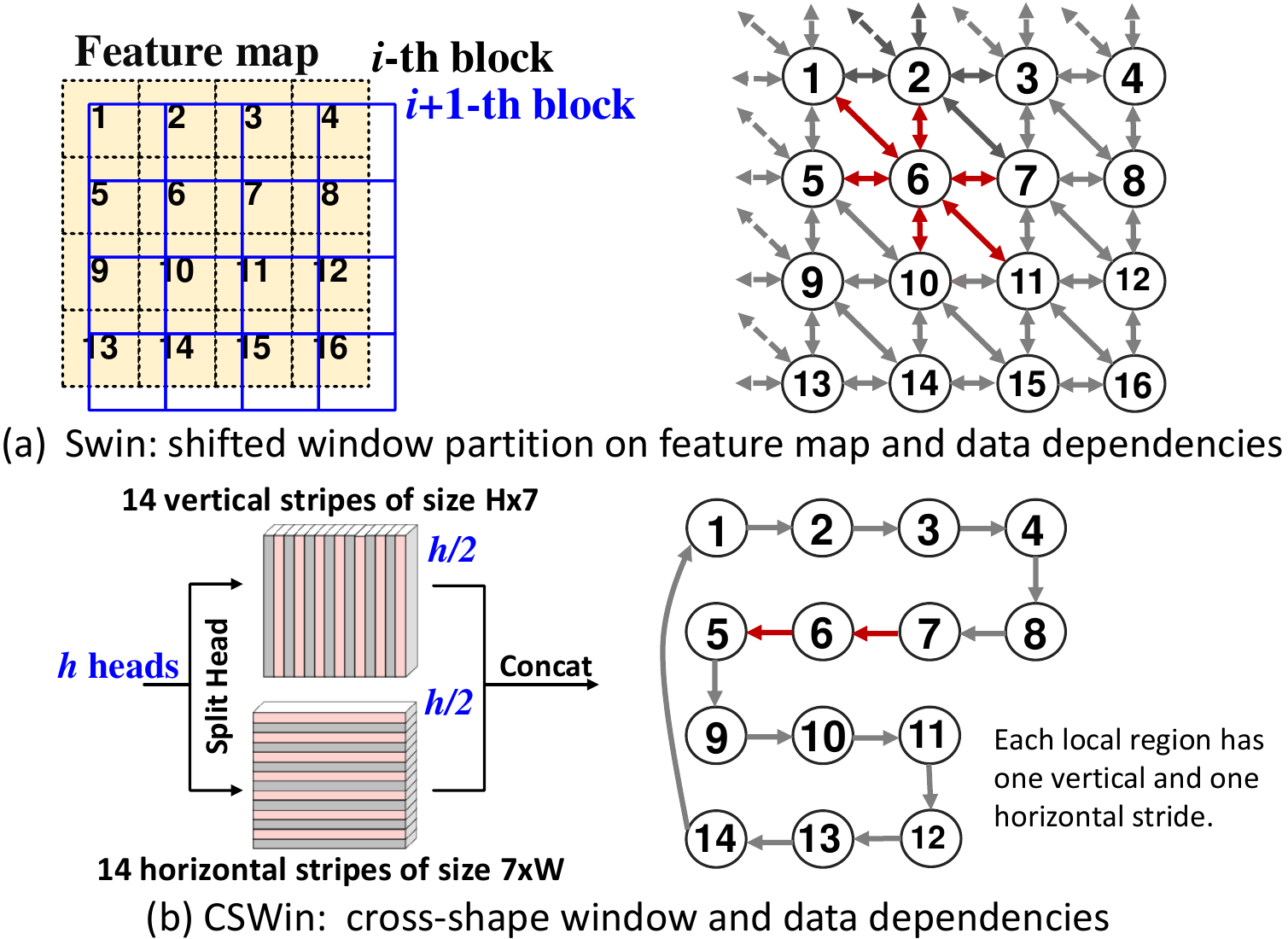}
\caption{Data dependencies between branches. Given an input of size $224^2$, the figures depict data dependencies between branches on encoder in 2nd model stage. Red edges mark the data dependency of 6th branch. As each PIM-node has to broadcast data to all remaining nodes (e.g., CSWin), to avoid cumbersome NoC communication, we employ cyclic data transfers.}
 \label{fig:data-depend}
\end{figure}

\subsection{Data Dependency for Local-Global Interaction}
As demonstrated in Section \ref{sec:local-global}, LVTs employ various complex operations to achieve local-global interaction within or between encoders in the same model stage.
Despite their complexity, these operations exhibit regular patterns, where each computational branch exchanges data only with certain neighboring ones, allowing for regularized data dependencies.
We define a graph $G_{dp}(V_{bk},E_{bk})$ to represent the data dependencies of different models, as shown in Figure \ref{fig:data-depend}.

\subsection{Two-stage Placement Method}
\label{sec:placement}
\begin{figure}[!t]
\centering
\includegraphics[width=0.5\textwidth]{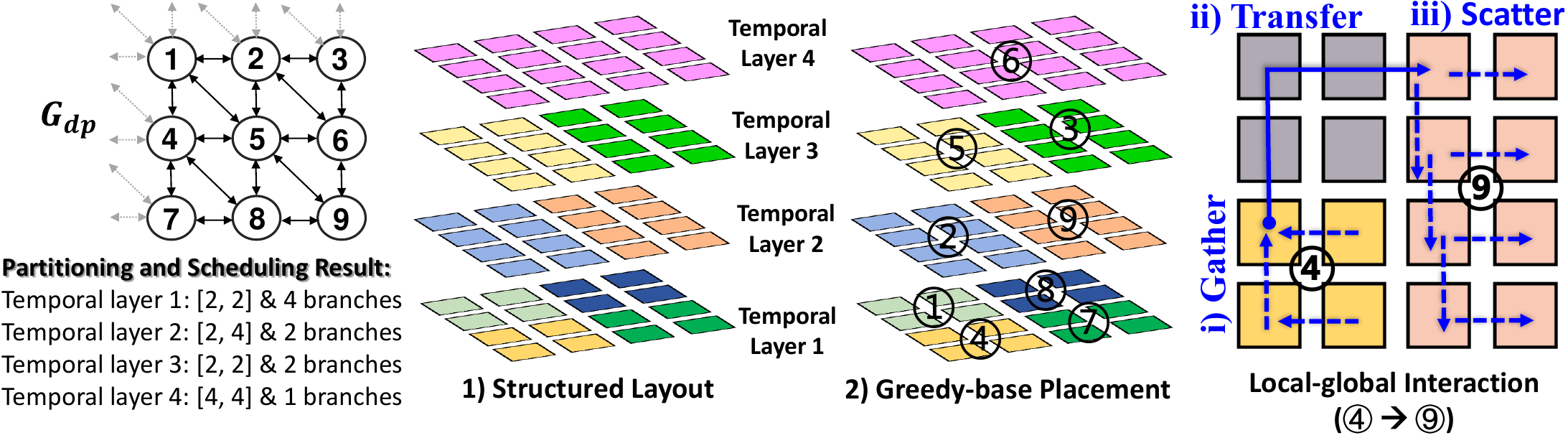}
\caption{Local-global interaction aware placement.}
\label{fig:mapping}
\end{figure}

The issue at hand involves placing all the computational branches of $|V_{bk}|$ local regions onto $n_{tl}$ temporal layers of the PIM system in a non-overlapping manner. This falls under the category of 3D placement in VLSI design as well as being NP-hard.
Importantly, all node sub-arrays occupied by computational branches on each temporal layer are \textbf{uniformly sized}.
Also, all computational branches within each encoder have exactly the same amount of computation, which facilitates an orderly layout.
In view of the orderly layout and regular data dependencies, we propose a two-stage heuristic of layout followed by binding, as shown in Figure \ref{fig:mapping}.

\subsubsection{Structured layout}
\label{sec:layout}
To ensure a maximum PIM-node occupancy, we systematically arrange $n_{br}$ uniformly-sized sub-arrays for each temporal layer of the PIM system of size $H_A \times W_A$ in Algorithm \ref{algo:layout}.
Note that, temporal layers with the same partition scheme adopt the $same$ su-barray layout.
The following example illustrates the structural layout when $H_A=3$, $W_A=4$, $u=2$, and $v=1$ with up to 6 sub-arrays:
\[\small\begin{bmatrix}
\varnothing & \varnothing & \varnothing & \varnothing\\
\varnothing & \varnothing & \varnothing & \varnothing\\
\varnothing& \varnothing & \varnothing & \varnothing
\end{bmatrix}
\xrightarrow[]{}
\begin{bmatrix}
1 & 2 & 3 & 4\\
1 & 2 & 3 & 4\\
\varnothing & \varnothing & \varnothing & \varnothing
\end{bmatrix}
\xrightarrow[\&Place]{Rotate}
\begin{bmatrix}
1 & 2 & 3 & 4\\
1 & 2 & 3 & 4\\
5 & 5 & 6 & 6
\end{bmatrix}\]
where $\varnothing$ stands for an unoccupied PIM-node.

\begin{algorithm}
  \fontsize{8.5}{9.5}\selectfont
  \caption{Structured Layout}
  \begin{algorithmic}[1]
    \REQUIRE $H_A$, $W_A$, $u$, $v$
    \ENSURE Structured layout for a maximum number of sub-arrays.
     \STATE $n_{window}^x,\ \texttt{layout$^{x}$}=\texttt{StructLayout}(H_A, W_A, u, v)$
     \STATE $n_{window}^y,\ \texttt{layout$^{y}$}=\texttt{StructLayout}(H_A, W_A, v, u)$
     \RETURN $(n_{window}^x \geq n_{window}^y)?\ \texttt{layout$^{x}$}:\texttt{layout$^{y}$}$\\
    \vspace{0.6em}
    \textbf{Function} $\texttt{StructLayout}(A, B, a, b)$:
    \STATE \quad $x_1 = \lfloor A/a\rfloor$, $y_1 = \lfloor B/b\rfloor$, $n_1=x_1*y_1$;
    \STATE \quad Place the $a$-side along the $A$-side and the $b$-side along the $B$-side to form the \texttt{layout} of $x_1\times y_1$ grid.
    \STATE \quad \textbf{if} {$a>b$ :}
    \STATE \quad \quad $A = A-x_1*a$;\quad $x_2 = \lfloor A/b\rfloor$;\quad $y_2 = \lfloor B/a\rfloor$.
    \STATE \quad \textbf{else}
    \STATE \quad \quad $B = B-y_1*b$;\quad $x_2 = \lfloor B/a\rfloor$;\quad $y_2 = \lfloor A/b\rfloor$.
    \STATE \quad Rotate and then place the sub-array to refine the \texttt{layout}.
    \STATE \quad $n_2=x_2*y_2$;
    \STATE \quad \textbf{return} $n_1+n_2$ and $\texttt{layout}$
  \end{algorithmic}
  \label{algo:layout}
\end{algorithm}

\subsubsection{Greedy-based binding}
\label{sec:binding}
with the regularized data dependency graph and the well-structured sub-array layout for each encoder, we propose a greedy-based method to bind the branches to specific PIM-nodes.
Firstly, we select the first computational branch and then place it on the top-left sub-array of the first temporal layer.
Next, based on $G_{dp}$, we select the branch with the most dependencies on already mapped branches and then place it on the sub-array with the minimal distance among all temporal layers.
This process is repeated iteratively until all computational branches are mapped.
In Figure \ref{fig:mapping}, the feature map of each computational branch distributed across its sub-array is first \textbf{i)gathered} to the top-leftmost PIM-node, then \textbf{ii)transferred} to the top-leftmost PIM-node of the other sub-array, and finally \textbf{iii)scattered} to its PIM-nodes.
Therefore, the distance is computed as the Manhattan distance between the top-leftmost PIM-nodes of two sub-arrays.
The method has a time complexity of $O(n_{br}^2)$.

Table \ref{table:place} shows the comparison between our method and DDAM\cite{DDAM}, which utilizes an ILP-based DNN mapping approach with high complexity. Even though the runtime of our method is much lower than that of DDAM, the average hop count of NoC communication at all model stages is lower.
\vspace{-1.4em}
\begin{table}[htbp]
  \centering
  \fontsize{8}{9}\selectfont
  \caption{Comparison of average hop counts in two placement methods.}
  \label{table:place}
  \begin{tabular}{|c|c|c|c|c|}
    \hline
    \textbf{Method} & \textbf{Stage 1}& \textbf{Stage 2}& \textbf{Stage 3} & \textbf{Stage 4}\\
    \hline
    \textbf{DDAM\cite{DDAM}} & 10.33& 8.48 & 6.42& 9.33 \\
    \hline
    \textbf{Ours} & 2.59& 4.51 & 3.41& 8.48 \\
    \hline
  \end{tabular}
\end{table}

\section{Experiments}
\label{sec:exp}
\subsection{Experiment Setting}
\label{sec:setting}
Proposed Allspark is implemented using Python on a Linux server with an Intel Xeon Gold 6254 CPU@3.10GHz server.

\textbf{PIM architecture:}
we adopt emerging 3D-stacked DRAM\cite{3DDRAM,VLSI} as the substrate of our PIM architecture, which provides a high-density, high-energy-efficient PIM solution built with logic-to-DRAM hybrid bonding technology.
Furthermore, the processing engine is built with the widely used NVDLA-style architecture\cite{Simba,NVDLA}.
This configuration allows the processing engines to be positioned closer to memory, which markedly reduces latency and power consumption of memory accesses and achieves an impressive energy efficiency of 0.66 pJ/bit\cite{VLSI}.
The detailed configurations are shown in Table~\ref{table:parameters}.

\textbf{Simulation:}
we utilize a DNN accelerator evaluation tool, \texttt{Timeloop+Accelergy}\cite{Timeloop}, to obtain the computation, memory accesses, and energy consumption of NVDLA-style processing engines in the PIM system.
To simulate DRAM access within and between PIM-nodes and energy consumption, we employ cycle-accurate simulation tools:  \texttt{Ramulator-PIM}\cite{Ramulator} in conjunction with \texttt{BookSim2.0}\cite{booksim} and \texttt{DRAMPower}\cite{DRAMPower}.
Following the integer quantization methods \texttt{i-BERT}\cite{ibert} and \texttt{FQ-ViT}\cite{FQ-ViT}, we synthesize dedicated ACU of 8-bit integers based on 28nm process technology using commercial logic synthesis tools to obtain processing delay for nonlinear computations.
Note that the computation remains unchanged during the inference on the PIM system and the accuracy of the visual model is unaffected.
\vspace{-1em}
\begin{table}[h!]
  \centering
  \fontsize{7.5}{10}\selectfont
  \caption{Configuration of the DRAM-based PIM System.}
  \label{table:parameters}
  \begin{tabular}{|l|l|l|}
    \hline
    \textbf{Module} &\textbf{Parameters} & \textbf{Configuration}\\
    \hline
    \hline
     \multirow{7}{*}{Logic Die} &Technology \& Clock Frequency & 28nm \& 400MHz \\
     &PIM-node Array \& PE Array& $16\times 16$ \& $8\times 8$\\
     &SRAM Buffers & 48 KiB\\
     &Bank Count per Node & One or more\\
    & Router & Input-queued architecture \\
    & Routing Algorithm& Dimension-order\\
    & Flit width \& Energy& 64-bit \& 1.1 pJ/bit/hop\\
    \hline
    \multirow{5}{*}{DRAM Die}& Technology & 25nm \\
    & Bank Capacity \& Bandwidth & 8 MiB \& 128 bit \\
    & Energy & 0.66 pJ/bit \\
    & Timing parameter (tRP, tRCD, & \multirow{2}{*}{18, 18, 40, 180, 3904 (ns)} \\
    & tRAS, tRFC, tREFI)& \\
    \hline
  \end{tabular}
\end{table}

\textbf{Baselines:}
state-of-the-art accelerators use either \emph{\textbf{attention-head-level(AH)}} partitioning\cite{DOTA,ViTCoD} or \emph{\textbf{patch-level(P)}} partitioning\cite{TransPIM,MAT,Spatten,FACT} for parallelization.
We conduct comparative assessments against AH, P and \emph{\textbf{branch-level(B)}} partitioning on the PIM system for various models.

In addition, we compare the inference latency of visual Transformers on a DRAM-based PIM system with Allspark against Nvidia V100 GPU.
We use \texttt{PyTorch} to implement the inference of different models running on GPU,
and record latency using \texttt{CUDA Events} and measure power consumption using \texttt{nvidia-smi}.
To this end, we scale the size of the node arrays and PE arrays of PIM system to achieve a peak throughput (14.7 TOPS@INT8) equivalent to that of GPU (15.7 TFLOPS). The GPU and PIM have aggregated memory capacities of 32 GiB and 4.5 GiB, respectively, and peak bandwidths of 900GB/s and 3.35TB/s, respectively.

\textbf{Workloads:}
we evaluate deployment frameworks across ViT\cite{ViT}, PVT\cite{PVT}, and four state-of-the-art LVT models, namely Swin\cite{Swin}, Focal\cite{Focal}, Twins\cite{Twins}, and CSWin\cite{CSWin}, using ImageNet dataset.
Differently, PVTs are deployed onto a node array of size 8$\times$8 due to its low parallelism.
Each model comes in various sizes: tiny(T), small(S), base(B), large(L), and huge(H).
These models have parameters ranging from 20 to 700 million and computations of hundreds of GFLOPs.

\subsection{Effectiveness of Allspark}
\label{sec:effect}

\begin{figure}[!t]
\centering
\subfloat[]{\includegraphics[width=0.49\textwidth]{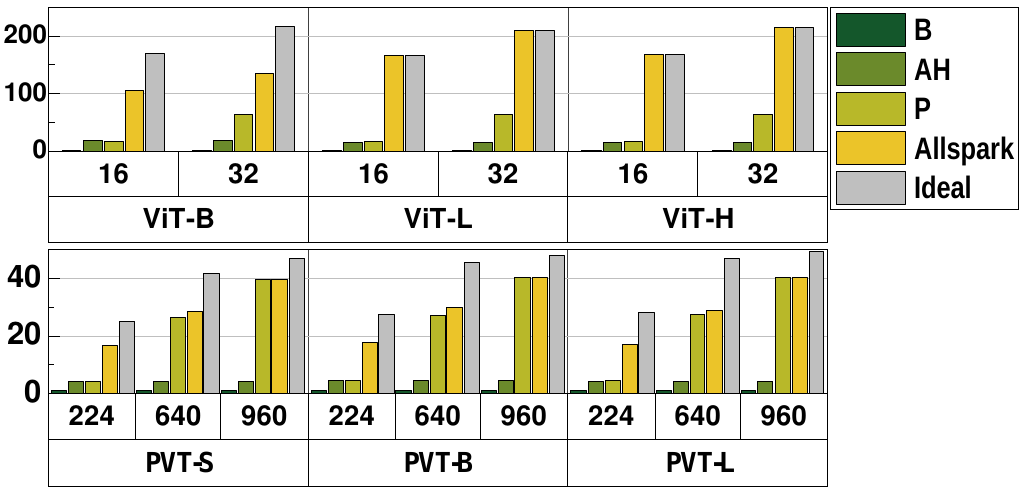}
\label{fig:mapper-1}}\\
\vspace{-0.8em}
\subfloat[]{\includegraphics[width=0.49\textwidth]{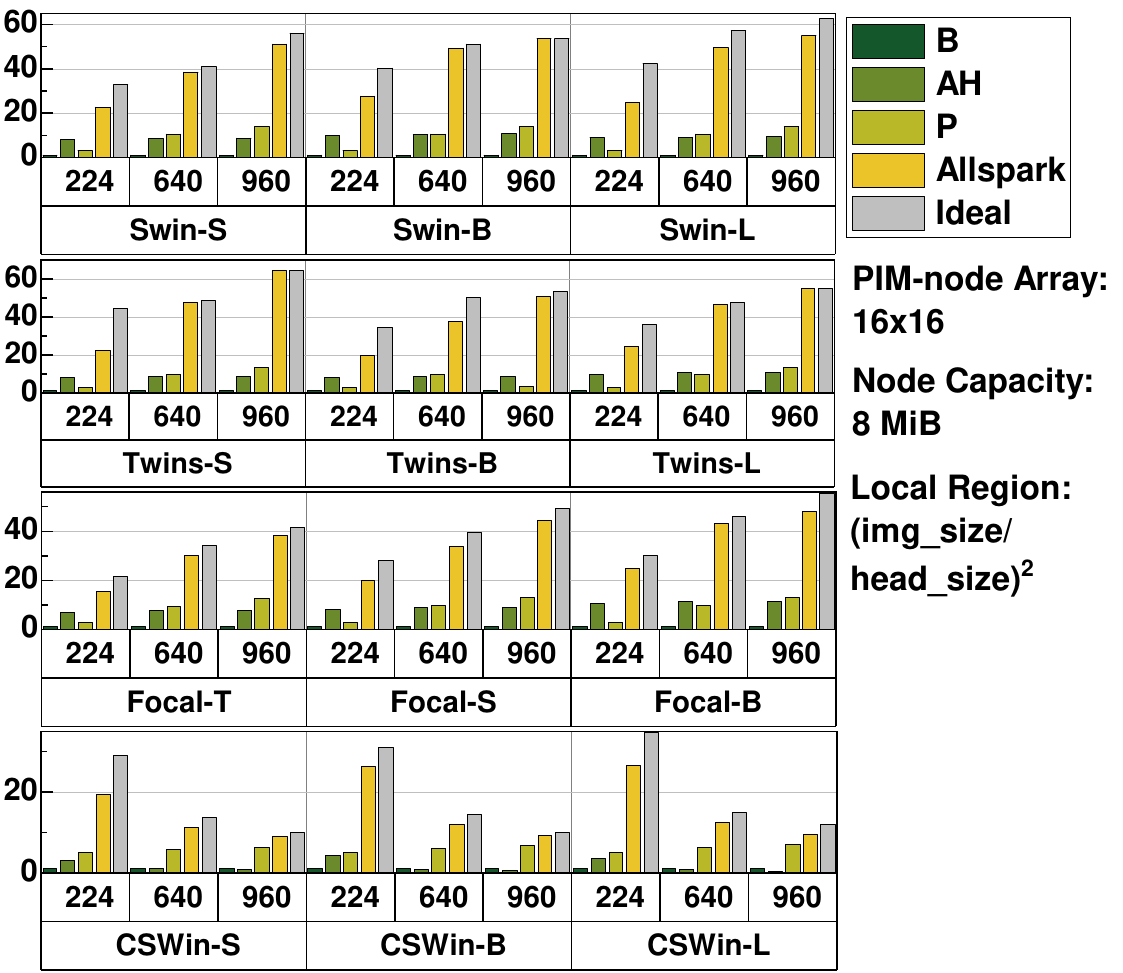}
\label{fig:mapper-2}}
\caption{Normalized speedups (w.r.t B) achieved by Allspark.}
\label{fig:mapper}
\end{figure}

The average searching overheads of Allspark are 0.8s(small models) and 3.4s(largest models) for input sizes of $224\times224$, increasing to 12.5min(small models) and 2.1h(largest models) for input sizes of $960\times960$, etc. For different models, the overhead difference is not significant. The Allspark optimization evaluation is within 10\% deviation from simulation results.

\textbf{1) Inference speed-up:}
As shown in Figure \ref{fig:mapper}, with the support of the proposed Allspark, the inference latency of models on the PIM system significantly improves.
For original visual Transformers, compared to B, P, and AH partitioning methods, the inference latency is improved on average by $28.8\times$, $6.7\times$, and $1.2\times$, respectively.
For LVTs, the average speedup is $40.0\times$, $5.1\times$, and $4.2\times$, respectively.
This improvement is particularly pronounced for larger input image sizes and larger models (except for CSWin).
In such cases, each transformer block has a greater number of attention heads and patches in the branch based on local regions, resulting in more partitioning possibilities.
Other methods use fixed and uni-patterned partitioning, while Allspark allows for fully flexible partitioning for each transformer block of models.
For larger input image and model sizes, CSWin Transformer has a higher number of branches (local regions) within its transformer block compared to other models.
Therefore, on CSWin, there is no significant speedup in Allspark compared to the branch-level method.
The figure also gives the ideal performance which assumes perfect hardware utilization.

\textbf{2) PIM-node utilization:}
We provide the PIM-node utilization for the staged and full stage of models, expressed as
\begin{equation}
\centering
\small
Node\_Util_s = \frac{\sum_{j}{N_{node\_occupied}^{s,j}\cdot Exec\_Time_{s,j}}}{\sum_{j}{(H_A\times W_A)\cdot Exec\_Time_{s,j}}}
\end{equation}
where $Exec\_Time_{s,j}$ and $N_{node\_occupied}^{s,j}$ are the runtime and the number of nodes occupied for the $j$-th temporal layer of the transformer block in the $s$-th model stage, respectively.

\begin{figure}[!t]
\centering
\includegraphics[width=0.46\textwidth]{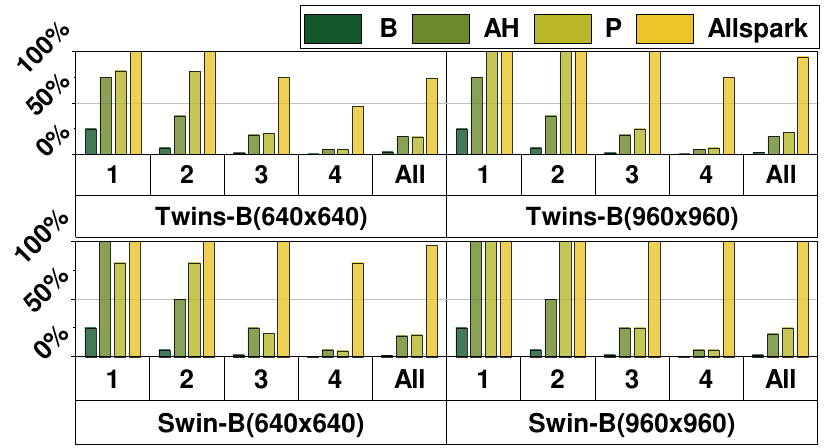}
\caption{Node utilization at each stage and all stages.}
\label{fig:node_util}
\end{figure}

Figure \ref{fig:node_util} shows node utilization of LVTs with different mapping methods at each stage and for all stages.
Using Allspark, PIM system has the highest node utilization in every stage of LVTs, and an overall utilization of over 96\% (Swin-B) and over 74\% (Twins-B), respectively.
In contrast, the node utilization of PIM system under other methods is below 20\%.
Overall, the node utilization tends to decrease from stage to stage due to the down-sampling performed in each stage, which causes a decreasing number of local regions (branches) in transformer blocks.
Furthermore, node utilization is dominated by the third model stage, which has the most transformer blocks (about 3 to 9 times more than other stages) and accounts for a large proportion of the computation.

\begin{figure}[!t]
\centering
\subfloat[Latency breakdown (cycle)]{\includegraphics[width=0.44\textwidth]{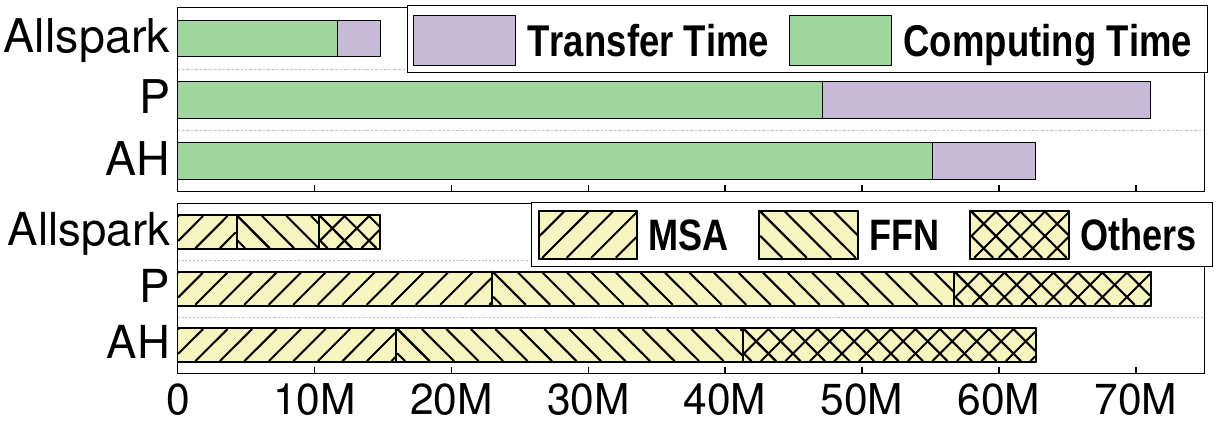}

    \label{fig:runtime-bd}}\\
\centering
\vspace{-1em}
\subfloat[Normalized comparison]{\includegraphics[width=0.44\textwidth]{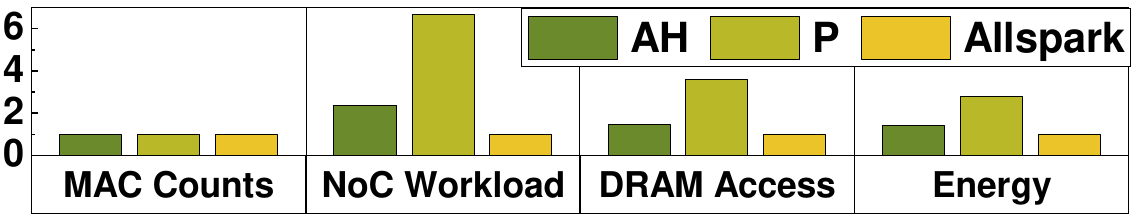}
    \label{fig:detail-bd}}
\caption{Inference latency breakdown and normalized comparison when Swin-B as well as the input image of size $640\times 640$.}
\label{fig:breakdown}
\end{figure}

\textbf{3) Latency breakdown:}
Figure \ref{fig:runtime-bd} reveals the inference latency breakdown, highlighting that Allspark enables LVTs to have minimal inference latency due to its high PIM-node utilization.
Furthermore, the time taken for feature map transfer and weight sharing is only 13\% to 41\% of theirs compared to other methods.
As the normalized comparison in Figure \ref{fig:detail-bd}, apart from the unchanged computation and MAC counts, Allspark enables PIM system to have less NoC workload, DRAM accesses, and energy consumption, with the patch-level partitioning approach showing the highest values, which are 6.7, 3.6, and 2.8 times higher than that of Allspark, respectively.
Those methods based on fixed and uni-pattern partitioning cause a heavy transmission for key-value ($KV$) matrices and massive weight sharing between PIM-nodes.

\begin{figure}[!t]
\centering
\includegraphics[width=0.47\textwidth]{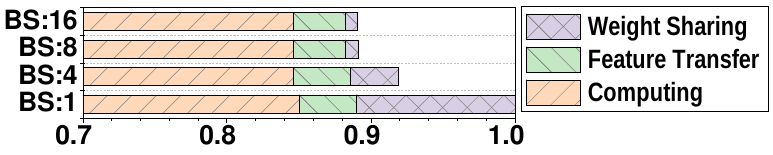}
\caption{Normalized average latency for small-batch inference.}
\label{fig:runtime-diffbatch}
\end{figure}

For different batch sizes(BS), Figure \ref{fig:runtime-diffbatch} shows the normalized average latency.
Overall, the computing time and feature map transfer time exhibit slight differences, while the average time taken for weight sharing decreases gradually as batch size increases, since weights are used in all batches after weight sharing at each computation phase, and when batch size is very large the weight sharing time is negligible.
As in Section \ref{sec:problem3}, the time spent on feature map transmission between transformer blocks (part of the feature map transfer in Figure \ref{fig:runtime-diffbatch}) is less than 1\%, confirming the soundness of prioritizing resource-driven partitioning and scheduling in Allspark.

\textbf{4) Memory usage:}
The solution provided by Allspark satisfies Constraint \ref{eq:cons-5}, where the memory capacity per PIM-node meets the space requirements during inference.
For Swin-B ($R_h$=$R_w$=7) and images of size $640\times 640$, with a memory capacity of 8 MiB per PIM-node, the solution for Allspark is in Table \ref{table:reuse}.
Table \ref{table:memory} provides an overview of memory usage.
After weight reuse and sharing, total memory usage drops to 7.96 MiB.
However, without either of them, memory required for each PIM-node would exceed its capacity (8 MiB).

\begin{table}[htbp]
  \centering
  \fontsize{8}{7.5}\selectfont
  \caption{Partitioning and Scheduling Result of Allspark.}
  \label{table:reuse}
  \begin{tabular}{cccc}
    \toprule
    \textbf{Stage} & \textbf{Temporal Layers}& \textbf{Partitioning Scheme$^\S$}& \textbf{Weight Reuse} \\
    \midrule
    1 & 3& \textcolor{blue}{[1,1], [1,1], [4,2]}&  \texttt{Yes}\\
    2 & 2& [2,1], [4,4]&  \texttt{No}\\
    3 & 3&  \textcolor{blue}{[8,2], [8,2],} [16,4]&   \texttt{Yes}\\
    4 & 1&  [4,4]&  \texttt{No}\\
    \bottomrule
  \end{tabular}
  \flushleft
  \ \ \footnotesize{$^\S$\textcolor{blue}{Blue text} indicates the temporal layer at which weight reuse occurs.}
\end{table}
\vspace{-1.2em}
\begin{table}[htbp]
  \centering
  \fontsize{8}{7.5}\selectfont
  \caption{Memory Requirements (MiB) for A Single Node.}%$^\ddag$.}
  \label{table:memory}
  \begin{tabular}{cccc}
    \toprule
    \textbf{Case} & \textbf{Weights}& \textbf{Workspace} & \textbf{Total} \\
    \midrule
    \textbf{w/ Reuse+Sharing} & 6.94& 1.02& 7.96 (\ding{51})\\
    w/o Reuse & 11.64& 1.02& 12.66 (\ding{55})\\
    w/o Sharing & 59.8& 1.02& 60.82 (\ding{55}) \\
    \bottomrule
  \end{tabular}
\end{table}

\begin{figure}[!t]
\centering
\includegraphics[width=0.45\textwidth]{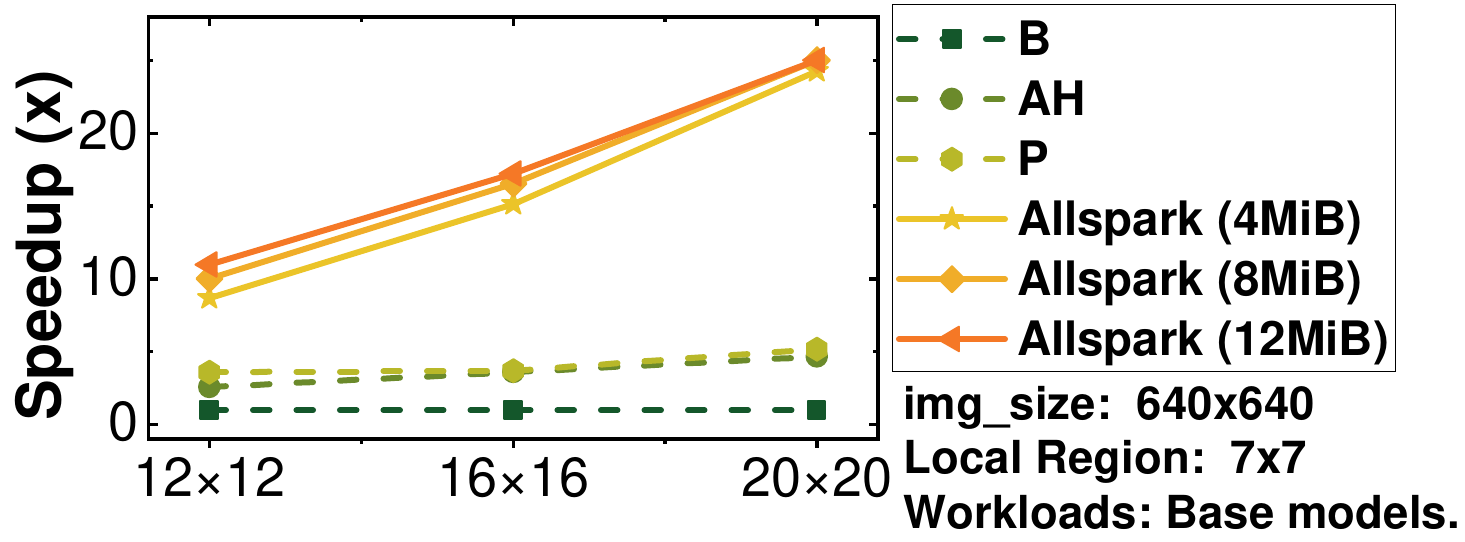}
\caption{Effectiveness of Allspark against the PIM-node array size and the memory capacity on each PIM-node.}
\label{fig:DSE}
\end{figure}

\textbf{5) Impact of PIM system configurations:}
The effectiveness of Allspark varies across PIM architectures, and Figure \ref{fig:DSE} shows the average normalized speedup for all base-size models.
As node array size and memory submodule scaling, Allspark has increasingly better performance over baselines due to its fully flexible partitioning and scheduling.
Memory submodule capacity per PIM-node is a hard constraint that governs the solution and affects the acceleration performance.

\subsection{Performance of Allspark-enriched PIM Systems}
\label{sec:GPU}
\begin{figure}[!t]
\centering
\includegraphics[width=0.49\textwidth]{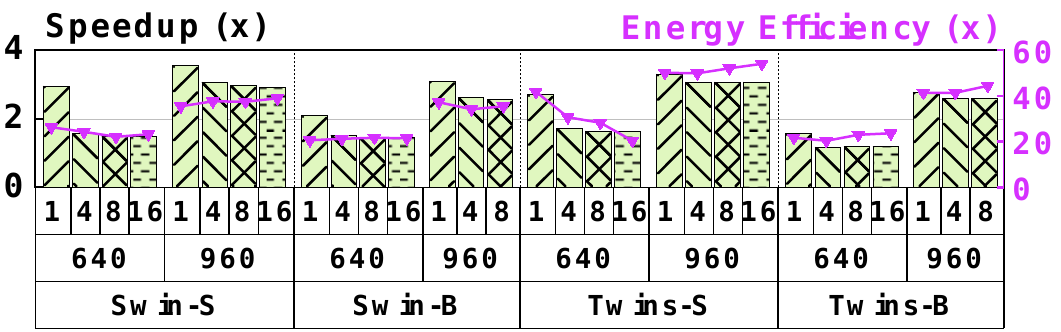}
\caption{Speed and energy efficiency improvements of DRAM-based PIM systems for small-batch inference over V100 GPU.}
\label{fig:comp-gpu}
\end{figure}

Normalized speedup and energy efficiency for small-batch inference on a 3D-stacked DRAM-based PIM system with Allspark and Nvidia V100 GPU are shown in Figure \ref{fig:comp-gpu}.
Allspark-enriched PIM systems have on average 2.3$\times$ speedup in latency as well as an improvement of $20\times$$\sim$$55\times$ in energy efficiency compared to GPUs for different models, model sizes, and batch sizes.
Especially for non-batched inference, the PIM system delivers the most significant speed gains, up to 3.5$\times$, and the speedups stabilize as batch size increases.
For the same model, the PIM system exhibits greater speedup when processing larger image sizes, attributed to a higher utilization of PIM-nodes, as in Figure \ref{fig:node_util}.
The difference between small and base models lies in the channel size, resulting in no significant variation in speedup.
Compared with other methods, Allspark brings over 2$\times$ speedup, exploiting the strengths of PIM system to achieve a high node utilization, due to its flexible partitioning strategy and scheduling scheme.

Continuously, we shrink the aggregate memory bandwidth of PIM to be the same as that of GPUs, at which point the PIM system has a smaller node array but a larger PE array per node. As shown in Table \ref{table:pim}, the Allspark-enriched PIM system still earns 1.95$\times$ average inference speedup.

\begin{table}[htbp]
  \centering
  \fontsize{7.8}{10}\selectfont
  \caption{Performance gains with scaled bandwidth.}
  \label{table:pim}
  \begin{tabular}{|c|c|c|c|c|}
    \hline
    \textbf{Memory Bandwidth} & \textbf{900 GB/s}& \textbf{1.12 TB/s}& \textbf{1.49 TB/s} & \textbf{3.35 TB/s} \\
    \hline
    \textbf{Average speedup} & 1.95& 2.15 & 2.30& 2.30\\
    \hline
  \end{tabular}
\end{table}

\section{Conclusions}
\label{sec:conclusion}
Allspark endeavors to workload orchestration for visual Transformers on the PIM systems, with the goal of minimizing inference latency.
Against the distributed nature of PIM system, Allspark endows a flexible partitioning strategy and elegant dataflows, partitioning and scheduling for end-to-end execution, and interaction-oriented placement, so that the system acquires high computational node utilization, reasonable weight arrangement, and efficacious data communication.
Evaluations on 3D-stacked DRAM-based PIM systems across various visual Transformers show that Allspark delivers significant inference speedups over baselines and Allspark-enriched PIM system yields notable improvements in both inference latency and energy efficiency over Nvidia V100 GPUs.
Visual Transformer models continue to evolve and spawn variants, yet the encoder module remains a pivotal component with significant computational and memory demands, which remains a deployment bottleneck. Consequently, Allspark continues to be available for the state-of-the-art visual Transformer models and variants, and it offers fertile ground for further enhancements.

\bibliographystyle{IEEEtran}
\bibliography{refs}

\begin{thebibliography}{10}
\providecommand{\url}[1]{#1}
\csname url@samestyle\endcsname
\providecommand{\newblock}{\relax}
\providecommand{\bibinfo}[2]{#2}
\providecommand{\BIBentrySTDinterwordspacing}{\spaceskip=0pt\relax}
\providecommand{\BIBentryALTinterwordstretchfactor}{4}
\providecommand{\BIBentryALTinterwordspacing}{\spaceskip=\fontdimen2\font plus
\BIBentryALTinterwordstretchfactor\fontdimen3\font minus
  \fontdimen4\font\relax}
\providecommand{\BIBforeignlanguage}[2]{{%
\expandafter\ifx\csname l@#1\endcsname\relax
\typeout{** WARNING: IEEEtran.bst: No hyphenation pattern has been}%
\typeout{** loaded for the language `#1'. Using the pattern for}%
\typeout{** the default language instead.}%
\else
\language=\csname l@#1\endcsname
\fi
#2}}
\providecommand{\BIBdecl}{\relax}
\BIBdecl

\bibitem{Attention}
A.~Vaswani, N.~Shazeer, N.~Parmar, J.~Uszkoreit, L.~Jones, A.~N. Gomez,
  L.~Kaiser, and I.~Polosukhin, ``Attention is all you need,'' in
  \emph{Proceedings of the 31st International Conference on Neural Information
  Processing Systems (NeurIPS)}, 2017, pp. 6000--6010.

\bibitem{TPAMI}
K.~Han, Y.~Wang, H.~Chen, X.~Chen, J.~Guo, Z.~Liu, Y.~Tang, A.~Xiao, C.~Xu,
  Y.~Xu, Z.~Yang, Y.~Zhang, and D.~Tao, ``A survey on vision transformer,''
  \emph{IEEE Transactions on Pattern Analysis and Machine Intelligence
  (TPAMI)}, vol.~45, no.~1, pp. 87--110, 2023.

\bibitem{ViT}
A.~Dosovitskiy, L.~Beyer, A.~Kolesnikov, D.~Weissenborn, X.~Zhai,
  T.~Unterthiner, M.~Dehghani, M.~Minderer, G.~Heigold, S.~Gelly, J.~Uszkoreit,
  and N.~Houlsby, ``An image is worth 16x16 words: Transformers for image
  recognition at scale,'' in \emph{International Conference on Learning
  Representations (ICLR)}, 2021.

\bibitem{PVT}
W.~Wang, E.~Xie, X.~Li, D.-P. Fan, K.~Song, D.~Liang, T.~Lu, P.~Luo, and
  L.~Shao, ``{Pyramid Vision Transformer}: A versatile backbone for dense
  prediction without convolutions,'' in \emph{2021 IEEE/CVF International
  Conference on Computer Vision (ICCV)}, 2021, pp. 548--558.

\bibitem{Swin}
Z.~Liu, Y.~Lin, Y.~Cao, H.~Hu, Y.~Wei, Z.~Zhang, S.~Lin, and B.~Guo, ``{Swin
  Transformer}: Hierarchical vision transformer using shifted windows,'' in
  \emph{Proceedings of the IEEE/CVF International Conference on Computer Vision
  (ICCV)}, October 2021, pp. 10\,012--10\,022.

\bibitem{Focal}
J.~Yang, C.~Li, P.~Zhang, X.~Dai, B.~Xiao, L.~Yuan, and J.~Gao, ``Focal
  attention for long-range interactions in vision transformers,'' in
  \emph{Proceedings of the 35th Conference on Neural Information Processing
  Systems (NeurIPS)}, vol.~34, 2021, pp. 30\,008--30\,022.

\bibitem{Twins}
X.~Chu, Z.~Tian, Y.~Wang, B.~Zhang, H.~Ren, X.~Wei, H.~Xia, and C.~Shen,
  ``{Twins}: revisiting the design of spatial attention in vision
  transformers,'' in \emph{Proceedings of the 35th International Conference on
  Neural Information Processing Systems}, 2024, pp. 9355--9366.

\bibitem{shuffle}
Z.~Huang, Y.~Ben, G.~Luo, P.~Cheng, G.~Yu, and B.~Fu, ``{Shuffle Transformer}:
  Rethinking spatial shuffle for vision transformer,'' \emph{arXiv preprint
  arXiv:2106.03650}, 2021.

\bibitem{CSWin}
X.~Dong, J.~Bao, D.~Chen, W.~Zhang, N.~Yu, L.~Yuan, D.~Chen, and B.~Guo,
  ``{CSWin Transformer}: A general vision transformer backbone with
  cross-shaped windows,'' in \emph{Proceedings of the IEEE/CVF Conference on
  Computer Vision and Pattern Recognition (CVPR)}, June 2022, pp.
  12\,124--12\,134.

\bibitem{sota}
\BIBentryALTinterwordspacing
{Papers with Code}: Latest papers with code. [Online]. Available:
  \url{https://paperswithcode.com/sota}
\BIBentrySTDinterwordspacing

\bibitem{TransPIM}
M.~Zhou, W.~Xu, J.~Kang, and T.~Rosing, ``{TransPIM}: A memory-based
  acceleration via software-hardware co-design for transformer,'' in \emph{2022
  IEEE International Symposium on High-Performance Computer Architecture
  (HPCA)}, 2022, pp. 1071--1085.

\bibitem{FACT}
Y.~Qin, Y.~Wang, D.~Deng, Z.~Zhao, X.~Yang, L.~Liu, S.~Wei, Y.~Hu, and S.~Yin,
  ``{FACT}: Ffn-attention co-optimized transformer architecture with eager
  correlation prediction,'' in \emph{Proceedings of the 50th Annual
  International Symposium on Computer Architecture}, 2023, pp. 1--14.

\bibitem{ViTCoD}
H.~You, Z.~Sun, H.~Shi, Z.~Yu, Y.~Zhao, Y.~Zhang, C.~Li, B.~Li, and Y.~Lin,
  ``{ViTCoD}: Vision transformer acceleration via dedicated algorithm and
  accelerator co-design,'' in \emph{2023 IEEE International Symposium on
  High-Performance Computer Architecture (HPCA)}, 2023, pp. 273--286.

\bibitem{Spatten}
H.~Wang, Z.~Zhang, and S.~Han, ``{SpAtten}: Efficient sparse attention
  architecture with cascade token and head pruning,'' in \emph{2021 IEEE
  International Symposium on High-Performance Computer Architecture (HPCA)},
  2021, pp. 97--110.

\bibitem{DOTA}
Z.~Qu, L.~Liu, F.~Tu, Z.~Chen, Y.~Ding, and Y.~Xie, ``{DOTA}: Detect and omit
  weak attentions for scalable transformer acceleration,'' in \emph{Proceedings
  of the 27th ACM International Conference on Architectural Support for
  Programming Languages and Operating Systems (ASPLOS)}, 2022, pp. 14--26.

\bibitem{GraphH}
G.~Dai, T.~Huang, Y.~Chi, J.~Zhao, G.~Sun, Y.~Liu, Y.~Wang, Y.~Xie, and
  H.~Yang, ``{GraphH}: A processing-in-memory architecture for large-scale
  graph processing,'' \emph{IEEE Transactions on Computer-Aided Design of
  Integrated Circuits and Systems}, vol.~38, no.~4, pp. 640--653, 2019.

\bibitem{Neurocube}
D.~Kim, J.~Kung, S.~Chai, S.~Yalamanchili, and S.~Mukhopadhyay, ``{Neurocube}:
  A programmable digital neuromorphic architecture with high-density 3d
  memory,'' in \emph{2016 ACM/IEEE 43rd Annual International Symposium on
  Computer Architecture (ISCA)}, 2016, pp. 380--392.

\bibitem{Newton}
M.~He, C.~Song, I.~Kim, C.~Jeong, S.~Kim, I.~Park, M.~Thottethodi, and T.~N.
  Vijaykumar, ``{Newton}: A dram-maker's accelerator-in-memory (aim)
  architecture for machine learning,'' in \emph{2020 53rd Annual IEEE/ACM
  International Symposium on Microarchitecture (MICRO)}, 2020, pp. 372--385.

\bibitem{VLSI}
S.~Wang, B.~Yu, W.~Xiao, F.~Bai, X.~Long \emph{et~al.}, ``A 135 gbps/gbit 0.66
  pj/bit stacked embedded dram with multilayer arrays by fine pitch hybrid
  bonding and mini-tsv,'' in \emph{2023 IEEE Symposium on VLSI Technology and
  Circuits}, 2023, pp. 1--2.

\bibitem{TETRIS}
M.~Gao, J.~Pu, X.~Yang, M.~Horowitz, and C.~Kozyrakis, ``{TETRIS}: Scalable and
  efficient neural network acceleration with 3d memory,'' in \emph{Proceedings
  of the Twenty-Second International Conference on Architectural Support for
  Programming Languages and Operating Systems (ASPLOS)}, 2017, pp. 751--764.

\bibitem{AiM}
S.~Lee, K.~Kim, S.~Oh, J.~Park, G.~Hong, D.~Ka, K.~Hwang, J.~Park, K.~Kang
  \emph{et~al.}, ``A 1ynm 1.25v 8gb, 16gb/s/pin gddr6-based
  accelerator-in-memory supporting 1tflops mac operation and various activation
  functions for deep-learning applications,'' in \emph{2022 IEEE International
  Solid- State Circuits Conference (ISSCC)}, vol.~65, 2022, pp. 1--3.

\bibitem{3DDRAM}
D.~Niu, S.~Li, Y.~Wang, W.~Han \emph{et~al.}, ``184qps/w 64mb/mm2 3d
  logic-to-dram hybrid bonding with process-near-memory engine for
  recommendation system,'' in \emph{2022 IEEE International Solid-State
  Circuits Conference (ISSCC)}, vol.~65, 2022, pp. 1--3.

\bibitem{DOJO}
E.~Talpes, D.~Williams, and D.~D. Sarma, ``{DOJO}: The microarchitecture of
  tesla's exa-scale computer,'' in \emph{2022 IEEE Hot Chips 34 Symposium
  (HCS)}, 2022, pp. 1--28.

\bibitem{nicepim}
J.~Wang, M.~Ge, B.~Ding, Q.~Xu, S.~Chen, and Y.~Kang, ``{NicePIM}: Design space
  exploration for processing-in-memory dnn accelerators with 3d-stacked-dram,''
  \emph{IEEE Transactions on Computer-Aided Design of Integrated Circuits and
  Systems}, vol.~43, no.~5, pp. 1456--1469, 2024.

\bibitem{MAT}
M.~Zhou, Y.~Guo, W.~Xu, B.~Li, K.~W. Eliceiri, and T.~Rosing, ``{MAT}:
  Processing in-memory acceleration for long-sequence attention,'' in
  \emph{2021 58th ACM/IEEE Design Automation Conference}, 2021, pp. 25--30.

\bibitem{TANGRAM}
M.~Gao, X.~Yang, J.~Pu, M.~Horowitz, and C.~Kozyrakis, ``{TANGRAM}: Optimized
  coarse-grained dataflow for scalable nn accelerators,'' in \emph{Proceedings
  of the Twenty-Fourth International Conference on Architectural Support for
  Programming Languages and Operating Systems (ASPLOS)}, 2019, pp. 807--820.

\bibitem{Timeloop}
A.~Parashar, P.~Raina, Y.~S. Shao, Y.-H. Chen, V.~A. Ying, A.~Mukkara,
  R.~Venkatesan, B.~Khailany, S.~W. Keckler, and J.~Emer, ``Timeloop: A
  systematic approach to dnn accelerator evaluation,'' in \emph{2019 IEEE
  International Symposium on Performance Analysis of Systems and Software
  (ISPASS)}, 2019, pp. 304--315.

\bibitem{Atomic}
S.~Zheng, X.~Zhang, L.~Liu, S.~Wei, and S.~Yin, ``Atomic dataflow based
  graph-level workload orchestration for scalable dnn accelerators,'' in
  \emph{2022 IEEE International Symposium on High-Performance Computer
  Architecture (HPCA)}, 2022, pp. 475--489.

\bibitem{CoSA}
Q.~Huang, M.~Kang, G.~Dinh, T.~Norell, A.~Kalaiah, J.~Demmel, J.~Wawrzynek, and
  Y.~S. Shao, ``{CoSA}: Scheduling by constrained optimization for spatial
  accelerators,'' in \emph{2021 ACM/IEEE 48th Annual International Symposium on
  Computer Architecture}, 2021, pp. 554--566.

\bibitem{inter-layer-scheduling}
J.~Cai, Y.~Wei, Z.~Wu, S.~Peng, and K.~Ma, ``Inter-layer scheduling space
  definition and exploration for tiled accelerators,'' in \emph{Proceedings of
  the 50th Annual International Symposium on Computer Architecture (ISCA)},
  2023.

\bibitem{DDAM}
J.~Wang, H.~Du, B.~Ding, Q.~Xu, S.~Chen, and Y.~Kang, ``{DDAM}: Data
  distribution-aware mapping of cnns on processing-in-memory systems,''
  \emph{ACM Transactions on Design Automation of Electronic Systems (TODAES)},
  vol.~28, no.~3, mar 2023.

\bibitem{Simba}
Y.~S. Shao, J.~Clemons, R.~Venkatesan, B.~Zimmer, M.~Fojtik, N.~Jiang,
  B.~Keller, A.~Klinefelter, N.~Pinckney, P.~Raina, S.~G. Tell, Y.~Zhang, W.~J.
  Dally, J.~Emer, C.~T. Gray, B.~Khailany, and S.~W. Keckler, ``{Simba}:
  Scaling deep-learning inference with multi-chip-module-based architecture,''
  in \emph{Proceedings of the 52nd Annual IEEE/ACM International Symposium on
  Microarchitecture (MICRO)}, 2019, pp. 14--27.

\bibitem{noc}
S.~M. Nabavinejad, M.~Baharloo, K.-C. Chen, M.~Palesi, T.~Kogel, and
  M.~Ebrahimi, ``An overview of efficient interconnection networks for deep
  neural network accelerators,'' \emph{IEEE JETCAS}, vol.~10, no.~3, pp.
  268--282, 2020.

\bibitem{TPU}
N.~P. Jouppi, C.~Young, N.~Patil, D.~Patterson, and \emph{et al},
  ``In-datacenter performance analysis of a tensor processing unit,'' in
  \emph{2017 ACM/IEEE 44th Annual International Symposium on Computer
  Architecture (ISCA)}, 2017, pp. 1--12.

\bibitem{NVDLA}
\BIBentryALTinterwordspacing
Nvidia, ``{NVDLA} deep learning accelerator,'' 2017. [Online]. Available:
  \url{http://nvdla.org.}
\BIBentrySTDinterwordspacing

\bibitem{BusMap}
X.~Ni, M.~Ge, Y.~Tao, W.~Sun, F.~Duan, X.~Bai, Q.~Xu, S.~Chen, and Y.~Kang,
  ``{BusMap}: Application mapping with bus routing for coarse-grained
  reconfigurable array,'' \emph{IEEE Transactions on Circuits and Systems II:
  Express Briefs}, vol.~70, no.~8, pp. 3054--3058, 2023.

\bibitem{FlexGen}
Y.~Sheng, L.~Zheng, B.~Yuan, Z.~Li, M.~Ryabinin, B.~Chen, P.~Liang, C.~Re,
  I.~Stoica, and C.~Zhang, ``{FlexGen}: High-throughput generative inference of
  large language models with a single {GPU},'' in \emph{Proceedings of the 40th
  International Conference on Machine Learning (ICML)}, vol. 202.\hskip 1em
  plus 0.5em minus 0.4em\relax PMLR, 23--29 Jul 2023, pp. 31\,094--31\,116.

\bibitem{SCALE-Sim}
A.~Samajdar, J.~M. Joseph, Y.~Zhu, P.~Whatmough, M.~Mattina, and T.~Krishna,
  ``A systematic methodology for characterizing scalability of dnn accelerators
  using scale-sim,'' in \emph{2020 IEEE International Symposium on Performance
  Analysis of Systems and Software (ISPASS)}, 2020, pp. 58--68.

\bibitem{ibert}
S.~Kim, A.~Gholami, Z.~Yao, M.~W. Mahoney, and K.~Keutzer, ``{I-BERT}:
  Integer-only bert quantization,'' \emph{arXiv preprint arXiv:2101.01321},
  2021.

\bibitem{FQ-ViT}
Y.~Lin, T.~Zhang, P.~Sun, Z.~Li, and S.~Zhou, ``{FQ-ViT}: Post-training
  quantization for fully quantized vision transformer,'' in \emph{Proceedings
  of the Thirty-First International Joint Conference on Artificial Intelligence
  (IJCAI)}, 7 2022, pp. 1173--1179.

\bibitem{gurobi}
\BIBentryALTinterwordspacing
G.~Optimization. (2023) Gurobi optimizer reference manual. [Online]. Available:
  \url{https://www.gurobi.com/}
\BIBentrySTDinterwordspacing

\bibitem{HPCC}
I.-H. Chung, C.-R. Lee, J.~Zhou, and Y.-C. Chung, ``Scalable
  communication-aware task mapping algorithms for interconnected multicore
  systems,'' in \emph{2011 IEEE International Conference on High Performance
  Computing and Communications}, 2011, pp. 759--764.

\bibitem{Ramulator}
Y.~Kim, W.~Yang, and O.~Mutlu, ``Ramulator: A fast and extensible dram
  simulator,'' \emph{IEEE Computer Architecture Letters}, vol.~15, no.~1, pp.
  45--49, 2016.

\bibitem{booksim}
N.~Jiang, D.~U. Becker, G.~Michelogiannakis, J.~Balfour, B.~Towles, D.~E. Shaw,
  J.~Kim, and W.~J. Dally, ``A detailed and flexible cycle-accurate
  network-on-chip simulator,'' in \emph{2013 IEEE International Symposium on
  Performance Analysis of Systems and Software}, 2013, pp. 86--96.

\bibitem{DRAMPower}
\BIBentryALTinterwordspacing
K.~Chandrasekar, C.~Weis, Y.~Li, S.~Goossens, M.~Jung, O.~Naji, B.~Akesson,
  N.~Wehn, and K.~Goossens, ``{DRAMPower}: Open-source {DRAM} power \& energy
  estimation tool,'' 2022. [Online]. Available: \url{http://www.drampower.info}
\BIBentrySTDinterwordspacing

\end{thebibliography}
% Photos in the template
\begin{IEEEbiography}[{\includegraphics[width=1in,height=1.25in,clip,keepaspectratio]{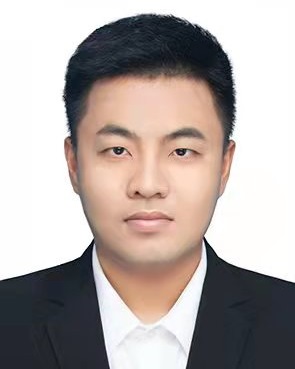}}]{Mengke Ge}
received the Ph.D. degree in electronic science and technology from the University of Science and Techonology of China (USTC), Hefei, China, in 2021.
He is currently an Associate Researcher with the Institute of Artificial Intelligence, Hefei Comprehensive National Science Center.
His research interests include AI-oriented compilation techniques, network-on-chip synthesis, and processing in-memory architecture.
\end{IEEEbiography}
\vspace{-2.5\baselineskip}
\begin{IEEEbiography}[{\includegraphics[width=1in,height=1.25in,clip,keepaspectratio]{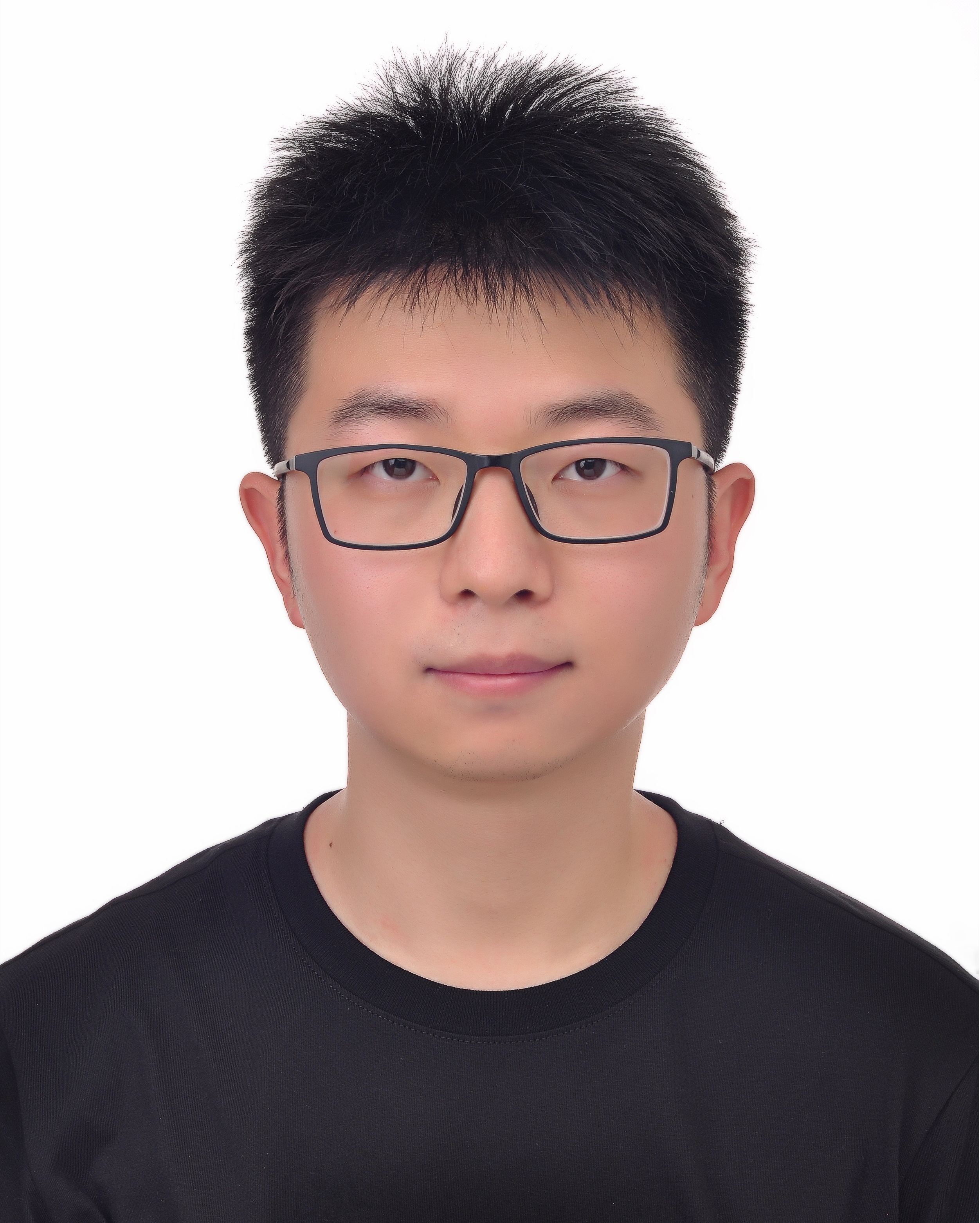}}]{Junpeng Wang}
received the B.S. degree in applied physics and the Ph.D. degree with the School of Microelectronics from USTC, Hefei, China, in 2018 and 2024, respectively.
His current research interests include hardware acceleration of deep neural networks and processing in-memory systems.
\end{IEEEbiography}
\vspace{-2.5\baselineskip}
\begin{IEEEbiography}[{\includegraphics[width=1in,height=1.25in,clip,keepaspectratio]{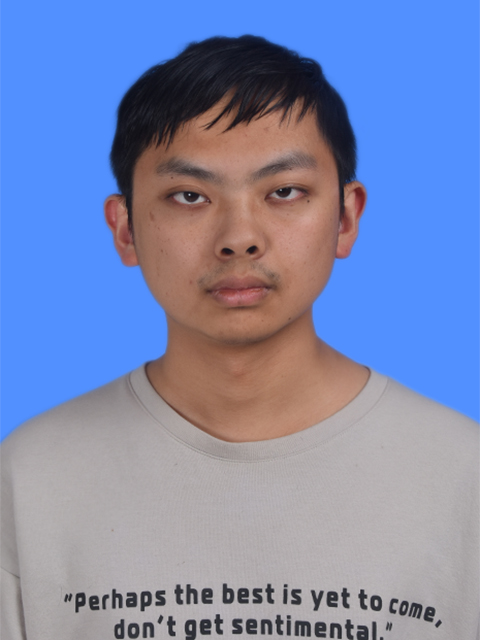}}]{Binhan Chen}
received the B.S. degree in electronic science and technology from USTC, Hefei, China, in 2020, where he is currently pursuing the Ph.D. degree with the School of Microelectronics. His current research interests include model pruning and model quantization.
\end{IEEEbiography}
\vspace{-2.5\baselineskip}
\begin{IEEEbiography}[{\includegraphics[width=1in,height=1.25in,clip,keepaspectratio]{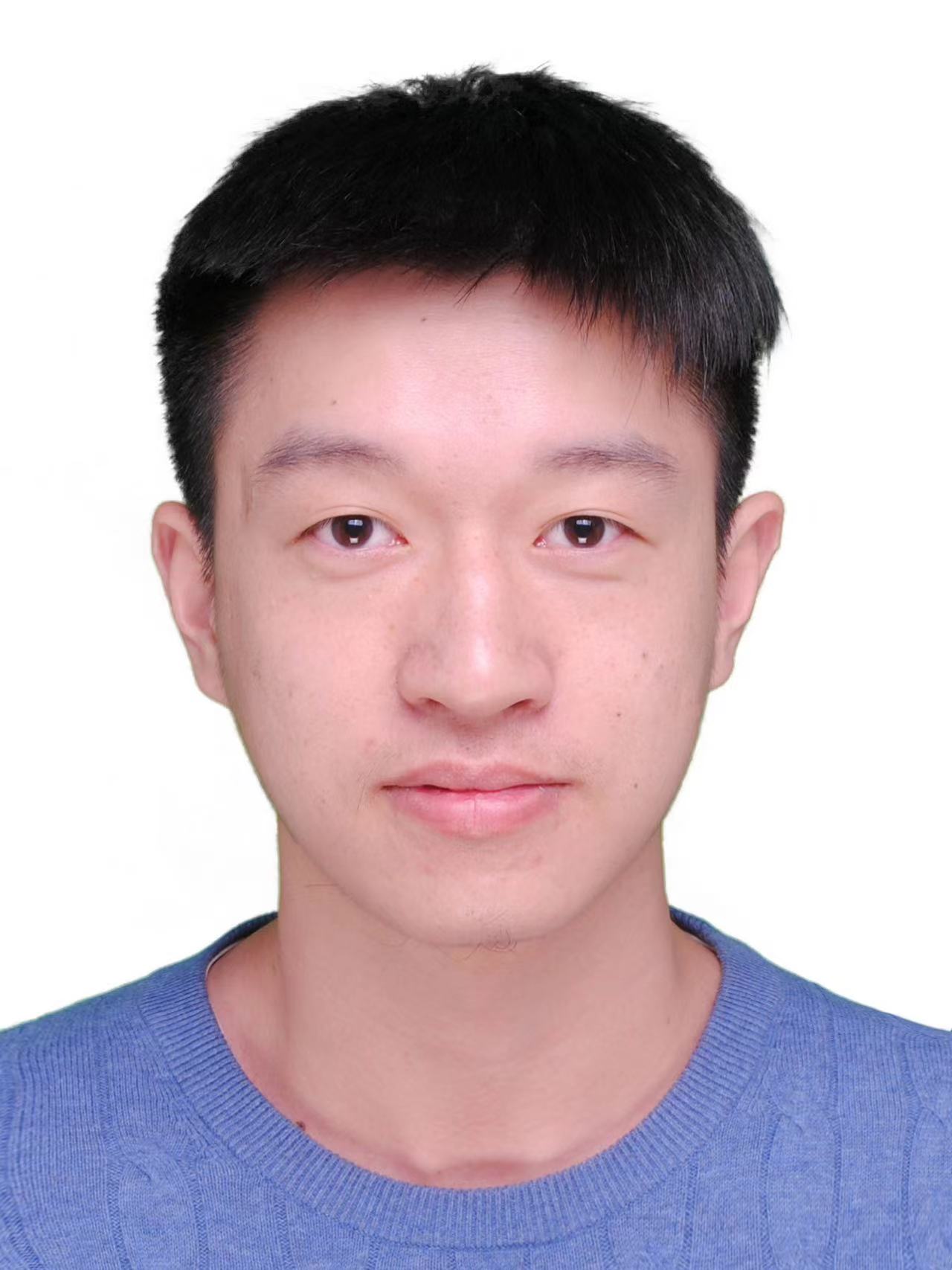}}]{Yingjian Zhong}
received the B.S. degree in Computer Science and Technology from Xidian University, Xi'an, China, in 2022, and is currently pursuing a MA.Eng degree in Artificial intelligence at Anhui University. His research interests include compilation for processing in-memory architectures.
\end{IEEEbiography}
\vspace{-2.5\baselineskip}
\begin{IEEEbiography}[{\includegraphics[width=1in,height=1.25in,clip,keepaspectratio]{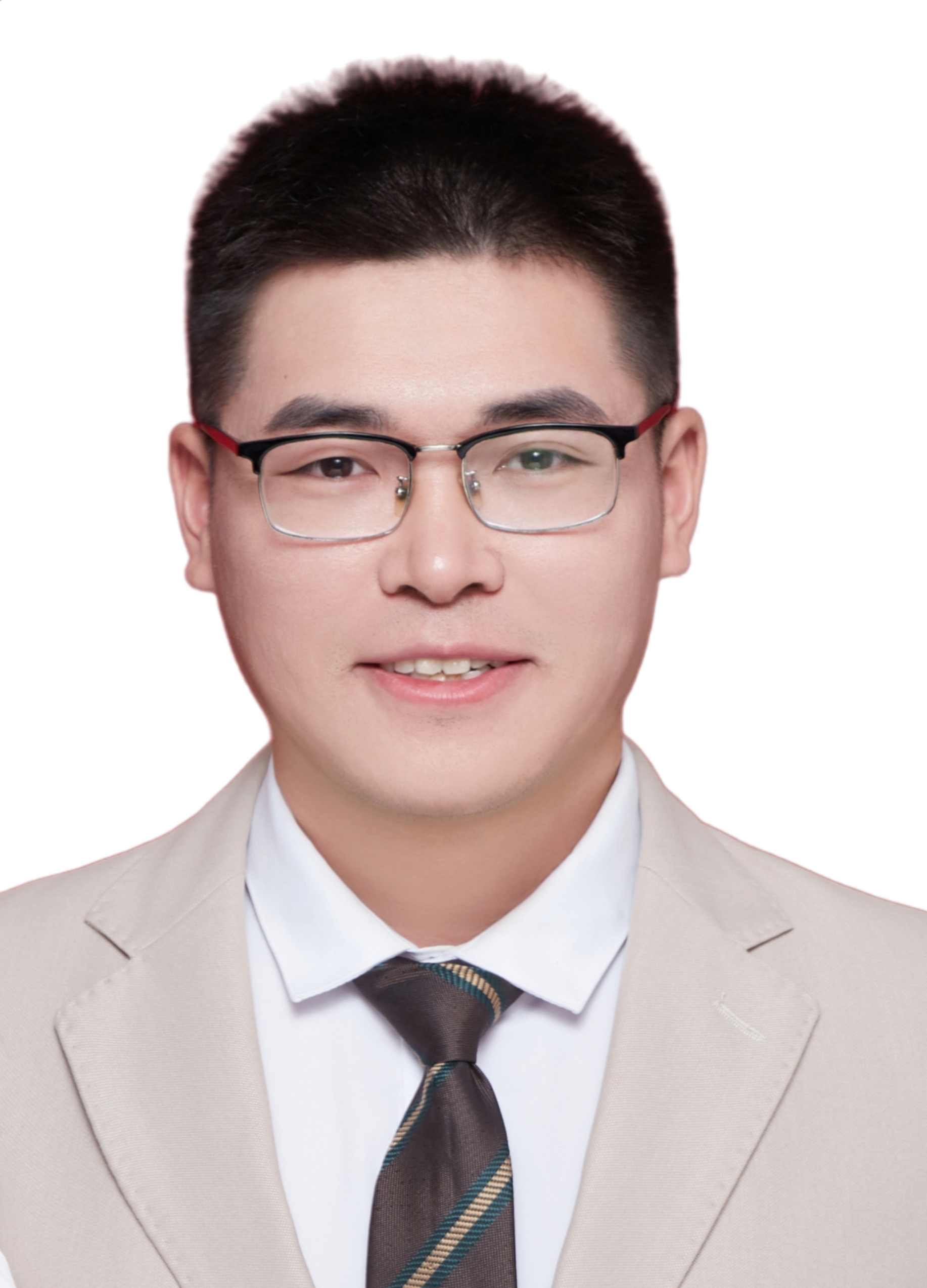}}]{Haitao Du}
is currently pursuing a Ph.D. degree at the School of Microelectronics, USTC in Hefei, China. His research interests primarily focus on DRAM memory architecture and processing in-memory systems.
\end{IEEEbiography}
\vspace{-2.5\baselineskip}
\begin{IEEEbiography}[{\includegraphics[width=1in,height=1.25in,clip,keepaspectratio]{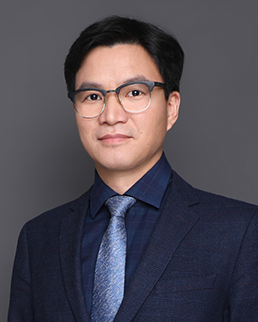}}]{Song Chen}
(Member, IEEE) received the B.S.degree in computer science from Xi'an Jiaotong University, China, in 2000, and the Ph.D. degree in computer science from Tsinghua University, China, in 2005. He served at the Graduate School of Information, Production and Systems, Waseda University, Japan, as a Research Associate from August 2005 to March 2009, and an Assistant Professor from April 2009 to August 2012. He is currently an Associate Professor with the School of Microelectronics, USTC. His research interests include several aspects of VLSI design automation, on-chip communication system, in-memory computing, and computer-aided design for emerging technologies. He is a member of ACM and IEICE.
\end{IEEEbiography}
\vspace{-2.5\baselineskip}
\begin{IEEEbiography}[{\includegraphics[width=1in,height=1.25in,clip,keepaspectratio]{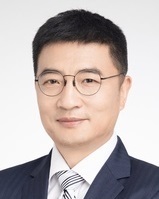}}]{Yi Kang}
(Member, IEEE) received the B.S. and M.S. degrees in electronic engineering from Tsinghua University and the Ph.D. degree in computer science from the University of Illinois at Urbana-Champaign.
He is currently a Professor with the School of Microelectronics, USTC.
Before went to teaching in USTC, he worked as the Chief Scientist and an SVP with Spreadtrum Communications Inc., Shanghai, China.
His current research area includes new computing and memory architecture and implementation of neural networks.
\end{IEEEbiography}

\end{document}